\documentclass[journal]{IEEEtran}

\usepackage{cite}
\usepackage{amsfonts,mathtools,bm,amssymb}
\usepackage{caption,subcaption}
\usepackage[inline]{enumitem}
\usepackage[dvipsnames]{xcolor}
\usepackage{cleveref}
\usepackage{algorithm}
\usepackage{algorithmicx}
\usepackage{algpseudocode}
\usepackage{xr}

\newtheorem{theorem}{Theorem}
\newtheorem{proposition}{Proposition}
\newtheorem{corollary}{Corollary}

\newtheorem{example}{Example}
\newtheorem{remark}{Remark}

\def\eqns#1{\begin{equation*}#1\end{equation*}}
\def\eqnsml#1{\begin{multline*}#1\end{multline*}}

\def\eqnl#1#2{\begin{equation}\label{#1}#2\end{equation}}
\def\eqnsa#1{\begin{align*}#1\end{align*}}

\def\one{\mathbf{1}}

\def\sfX{\mathsf{X}}
\def\sfY{\mathsf{Y}}
\def\sfZ{\mathsf{Z}}
\def\uvn{\bm{n}}
\def\uvm{\bm{m}}
\def\uvu{\bm{u}}
\def\uvv{\bm{v}}
\def\uvx{\bm{x}}
\def\uvy{\bm{y}}
\def\uvz{\bm{z}}

\def\calX{\mathcal{X}}
\def\calY{\mathcal{Y}}
\def\calZ{\mathcal{Z}}

\def\bbE{\mathbb{E}}
\def\bbI{\mathbb{I}}
\def\bbN{\mathbb{N}}
\def\bbP{\mathbb{P}}
\def\bbR{\mathbb{R}}
\def\bbV{\mathbb{V}}
\def\bbX{\mathbb{X}}

\def\bbZ{\mathbb{Z}}

\def\b{\mathrm{b}}
\def\c{\mathrm{c}}
\def\d{\mathrm{d}}
\def\p{\mathrm{p}}
\def\s{\mathrm{s}}

\def\dc{\mathrm{dc}}
\def\df{\mathrm{df}}
\def\ns{\mathrm{ns}}
\def\fa{\mathrm{fa}}

\def\Sym{\mathrm{Sym}}

\def\oN{\overline{\mathrm{N}}}
\def\N{\mathrm{N}}

\DeclareMathOperator*{\argmax}{argmax}

\def\given{\,|\,}
\def\Given{\,\big|\,}
\def\ind#1{\one_{#1}}

\def\tr{\intercal}
\def\ninf#1{\|#1\|_{\infty}}
\def\bscdot{\bm{\cdot}}

\begin{document}

\title{A linear algorithm for multi-target tracking \\ in the context of possibility theory}

\author{Jeremie~Houssineau\thanks{J.\ Houssineau is with the Department of Statistics at the University of Warwick.}}%

\maketitle

\begin{abstract}
We present a modelling framework for multi-target tracking based on possibility theory and illustrate its ability to account for the general lack of knowledge that the target-tracking practitioner must deal with when working with real data. We also introduce and study variants of the notions of point process and intensity function, which lead to the derivation of an analogue of the probability hypothesis density (PHD) filter. The gains provided by the considered modelling framework in terms of flexibility lead to the loss of some of the abilities that the PHD filter possesses; in particular the estimation of the number of targets by integration of the intensity function. Yet, the proposed recursion displays a number of advantages such as facilitating the introduction of observation-driven birth schemes and the modelling the absence of information on the initial number of targets in the scene. The performance of the proposed approach is demonstrated on simulated data.
\end{abstract}


\section{Introduction}

\IEEEPARstart{M}{ulti}-target tracking refers to the problem of estimating the states of an unknown number of dynamical targets based on point observations marred by uncertainty \cite{BarShalom1990,Stone2013}. The relationship between states and observations might be non-linear and some components of the state might be hidden in general. The targets are also subject to a birth-death process. The main difficulty lies in the fact that the target-originated observations are not labelled from one time step to the other and that they are mixed up with noise-originated observations called false alarms. This particular aspect of multi-target tracking, usually referred to as the data-association problem, is highly combinatorial in nature. The most natural model for multiple targets is to consider the credibility for a given sequence of observations to originate from a unique target. This sequence of observation, together with an hypothesised time of birth, corresponds to a potential target and is usually referred to as a track. Although the concept of track is useful, its introduction requires solving the data association problem explicitly \cite{Reid1979}. As a consequence, the computational complexity can only be reduced via approximations \cite{Fortmann1980}. Another solution is to give up the concept of track and focus instead on the population of targets as a whole. The appropriate mathematical concept in this case is the one of point process \cite{Daley2003, Streit2010}. Solutions to the multi-target tracking problem based on this concept can be traced back to \cite{Mori1986} and \cite{Washburn1987}, with \cite{Mori1986} using the corresponding random sets instead. Note that it is usual in the point-process literature to alternate between the two formalisms depending on the context \cite{Chiu2013}. The random set formulation subsequently became dominant in the field of multi-target tracking with the derivation of the probability hypothesis density (PHD) filter \cite{Mahler2003} and the introduction of its practical implementations in \cite{Vo2005} and \cite{Vo2006}. Other approaches to multi-target tracking include the use of convolutional neural networks \cite{Nam2016}, which are particularly suitable for visual tracking. At the level of a single track, alternatives to Bayesian inference include least squares polynomial fitting \cite{Li2018_joint} which allow for more modelling flexibility.

As for any complex system, it is challenging to characterise the targets' dynamics and observation, the birth-death process they are subject to and the false alarms. Defining statistical models for all uncertain aspects of the problem requires the introduction of many distributional assumptions together with the corresponding parameters. When dealing with real data, the target-tracking practitioner knows that his statistical model will be incorrect and can only hope to capture some aspects of the complex and varied processes at play. This inherent misspecification often implies that a number of heuristics have to be introduced to compensate for the discrepancies between the model and the data. It would therefore be beneficial to have a pragmatic approach where the modeller can easily acknowledge the limited amount of information that is available about the multi-target systems of interest without resorting to the use hyper-parameters. This is the motivation behind a number of theories such as possibility theory \cite{Zadeh1978, Dubois2015} and Dempster-Shafer theory \cite{Dempster1968, Shafer1976}, which aim at providing flexible representations of uncertainty. Dempster-Shafer theory is well-known in tracking, it is however usually applied on the output of probabilistic inference algorithms such as for data association \cite{Denoeux2014} or for target identification/classification \cite{Buede1997}. Alternative representations of uncertainty were also discussed in \cite{Mahler2007} for modelling different types of observed information, leading to a generalised form of likelihood function. We will focus on possibility theory since the standard probabilistic concepts are more easily extended to this context. Conventions for naming concepts and operations will be slightly different from standard possibility theory and will follow instead the approach of \cite{Houssineau2018_parameter} where possibility and probability theories are combined to form a general framework for statistical inference.

We will start by reviewing the necessary concepts and results in \Cref{sec:uncertain} before moving on to the introduction and study of an analogue of the notion of point process in \Cref{sec:uncertainCountingMeasure}. A complete multi-target model will then be defined in \Cref{sec:model}, followed by the introduction of a recursion akin to the PHD filter in \Cref{sec:recursion}. Simulation results are presented in Section~\ref{sec:simulations} before concluding in \Cref{sec:conclusion}.

\section{Uncertain variable}
\label{sec:uncertain}

The objective in this article will be to follow as closely as possible the standard probabilistic approach to multi-target modelling, but in the context of possibility theory. For this reason, we start by introducing an analogue of the notion of random variable as follows: let $\Omega$ be a sample space containing all the possible ``states of nature''. As opposed to the sample space used in probability theory, $\Omega$ is not equipped with a fundamental probability distribution (usually denoted $\bbP$) and is instead assumed to contain the true state of nature $\omega^*$. This construction highlights the nature of the approach: the quantities of interest are not random, they are simply unknown, and we aim to find their true values out of a set of possible values. Here, we use the word ``random'' in the strict sense, i.e.\ for an experiment that would yield given frequencies if repeated multiple times. We are not interested in learning the entirety of the true state of nature in general and focus instead on specific quantities such as the position of a given target. This can be formalised by introducing a function $\uvx : \Omega \to \sfX$ with $\sfX$ the space where the quantity of interest lives. The function $\uvx$ is referred to as a (deterministic) uncertain variable and will play the same role as random variables in probability theory. The true value of the quantity of interest is equal to $\uvx(\omega^*)$ by construction. Our current knowledge about $\uvx$ can be encoded in a function $f_{\uvx} : \sfX \to [0,1]$ such that $f_{\uvx}(x)$ is the credibility for the event $\uvx = x$. Importantly, $f_{\uvx}$ is not a density function even if $\sfX$ is uncountable, e.g.\ $\sfX = \bbR$, and the correct assumption about the normalisation of $f_{\uvx}$ is $\sup_{x \in \sfX} f_{\uvx}(x) = 1$ or $\max_{x \in \sfX} f_{\uvx}(x) = 1$ if $\sfX$ is countable, e.g.\ $\sfX = \bbN_0 = \{0, 1, 2, \dots\}$. This assumption is more flexible than the usual one for probability distributions where the supremum and maximum are replaced by integrals and sums; for instance, we can simply set $f_{\uvx}(x) = 1$ for any $x \in \sfX$ if nothing is known about $\uvx$, even if $\sfX$ is unbounded or infinite-dimensional. Uncertain variables do not induce a unique possibility function since the latter only quantifies what is known about $\uvx$ and we say that $f_{\uvx}$ \emph{describes} $\uvx$. The information about $\omega^*$ induced by $f_{\uvx}$ is represented by another possibility function $f$ on $\Omega$, defined as
\begin{equation}
\label{eq:fundamentalPossibilityFunction}
f(\omega) = f_{\uvx}(\uvx(\omega)), \qquad \omega \in \Omega.
\end{equation}

If $\uvy$ is another uncertain variables on $\sfY$ and if the possibility function $f_{\uvx,\uvy}$ on $\sfX \times \sfY$ describes $\uvx$ and $\uvy$ jointly, then the marginal and posterior possibility functions are \cite{Zadeh1978,deBaets1999}
\begin{subequations}
\begin{align}
\label{eq:marginal}
f_{\uvy}(y) & = \sup_{x \in \sfX} f_{\uvx,\uvy}(x, y) \\
\label{eq:BayesRule}
f_{\uvx|\uvy}(x \given y) & = \dfrac{f_{\uvx,\uvy}(x, y)}{f_{\uvy}(y)},
\end{align}
\end{subequations}
where it appears that analogues of standard probabilistic results often take a similar form but with supremums instead of integrals and possibility functions instead of probability density functions (p.d.f.s). Using these notions of marginal and conditional, \eqref{eq:BayesRule} can be expressed as
\begin{equation}
\label{eq:BayesRule2}
f_{\uvx|\uvy}(x \given y) = \dfrac{f_{\uvy|\uvx}(y \given x) f_{\uvx}(x)}{\sup_{x' \in \sfX} f_{\uvy|\uvx}(y \given x') f_{\uvx}(x')},
\end{equation}
where the form of Bayes' rule as used in statistical inference is easily recognisable. Similarly, if it holds that $f_{\uvx,\uvy}(x,y) = f_{\uvx}(x) f_{\uvy}(y)$ for all $(x,y) \in \sfX \times \sfY$ then $\uvx$ and $\uvy$ are said to be \emph{weakly independent} or, alternatively, \emph{independently described}. This notion of independence only implies that the information we have about $\uvx$ is not related to $\uvy$ and conversely. For instance, if we are told that two objects of interest are approximately 2 meters away from each other, then the information we hold about the first object is \emph{not} independent of the information we hold about the second object.

The marginalisation rule \eqref{eq:marginal} is a special case of the change of variable formula \cite{Baudrit2008}: let $\uvx$ and $\uvz$ be two uncertain variables  in $\sfX$ and $\sfZ$, respectively, verifying $\uvz = T(\uvx)$ for a given mapping $T : \sfX \to \sfZ$, if $\uvx$ is described by $f_{\uvx}$ then $\uvz$ is described by
\begin{equation}
\label{eq:changeOfVariable}
f_{\uvz}(z) = \sup_{x \in \sfX : T(x) = z} f_{\uvx}(x), \qquad z \in \sfZ.
\end{equation}
This change of variable formula does not contain a Jacobian term since possibility functions are not densities. One of the important consequences is that an uninformative possibility function for $\uvx$, i.e.\ $f_{\uvx}(x) = 1$ for any $x \in \sfX$, induces an uninformative possibility function for $\uvz$ in general. This is not the case with the uniform probability distribution which depends on the parametrisation.

In order to introduce meaningful notions of expected value and variance, a law of large numbers and a central limit theorem have been derived in \cite{Houssineau2019}. The law of large numbers yields the following definition of expected value:
$$
\bbE^*(\uvx) = \argmax_{x \in \sfX} f_{\uvx}(x),
$$
which identifies the expected value with the mode of the possibility function $f_{\uvx}$; we will assume that $\bbE^*(\uvx)$ is a singleton. This is consistent with the fact that we are interested in a single point, i.e.\ the true value $\uvx(\omega^*)$, and $\bbE^*(\uvx)$ is where this value is the most likely to be found. This notion of expected value verifies $\bbE^*(T(\uvx)) = T(\bbE^*(\uvx))$ for any mapping $T$ on $\sfX$ \cite{Houssineau2019}. Similarly, the central limit theorem yields the following notion of variance:
\begin{subequations}
\begin{align}
\label{eq:variance}
\bbV^*(\uvx) & = \bigg( -\dfrac{\d^2}{\d x^2} \log f_{\uvx}\big(\bbE^*(\uvx)\big) \bigg)^{-1} \\
&  = \bbE^*\bigg( -\dfrac{\d^2}{\d x^2} \log f_{\uvx}(\uvx) \bigg)^{-1}
\end{align}
\end{subequations}
where $f_{\uvx}$ is assumed to be twice differentiable at $\bbE^*(\uvx)$. This notion of variance can be seen as the inverse of a notion of Fisher information, which further justifies the interpretation of uncertain variables and possibility functions as representing (a lack of) information rather than randomness. The limiting possibility function in the central limit theorem is the Gaussian possibility function defined as
\begin{equation}
\label{eq:Gaussian}
\oN(x; \mu, \sigma^2) = \exp\Big(  -\dfrac{1}{2\sigma^2}(x- \mu)^2 \Big), \qquad x \in \bbR,
\end{equation}
with expected value $\mu \in \bbR$ and variance $\sigma^2 > 0$, which indicates that the Gaussian possibility function plays the same role in possibility theory as the Gaussian p.d.f.\ in probability theory. There is another notion of expectation, which is the direct analogue of corresponding notion for random variables, defined as
$$
\bar{\bbE}(\varphi(\uvx)) = \sup_{x \in \sfX} \varphi(x) f_{\uvx}(x),
$$
for any real-valued function $\varphi$ on $\sfX$. The scalar $\bar{\bbE}(\varphi(\uvx))$ can be interpreted as the maximum expected value of $\varphi(\uvx)$. When the argument of the supremum is non-negative, as in $\bar{\bbE}(\varphi(\uvx))$ when $\varphi$ is non-negative, the supremum can be identified with the uniform norm $\ninf{\bscdot}$, e.g.\ $\bar{\bbE}(\varphi(\uvx)) = \ninf{\varphi \cdot f_{\uvx}}$.

In probability theory, p.d.f.s are a simpler way of expressing probability measures. Similarly, a possibility function $f_{\uvx}$ is related to a more formal set function $\bar{P}_{\uvx}$ defined as
$$
\bar{P}_{\uvx}(B) = \sup_{x \in B} f_{\uvx}(x), \qquad B \subseteq \sfX.
$$
The set function $\bar{P}_{\uvx}$ satisfies most of the conditions for qualifying as a probability measure except additivity. Instead, $\bar{P}_{\uvx}$ is an outer measure verifying $\bar{P}_{\uvx}(\sfX) = 1$, so that we refer to it as an \emph{outer probability measure}. The scalar $\bar{P}_{\uvx}(B)$ is simply the credibility of the event $\uvx \in B$. This type of set function is called a plausibility measure in possibility theory and Dempter-Shafer theory, but the name outer probability measure will turn out to be more convenient when studying an analogue of the notion of point process in \Cref{sec:uncertainCountingMeasure}. One outer probability measure of particular importance is the one induced by the possibility function $f$ on $\Omega$ as defined in \eqref{eq:fundamentalPossibilityFunction}; we denote this outer probability measure by $\bar{\bbP}$. The credibility of an event $\uvx \in B$ can now be written $\bar{\bbP}(\uvx \in B)$ by identifying $\uvx \in B$ with the subset $\{\omega \in \Omega : \uvx(\omega) \in B\}$ of $\Omega$, as is usual in probability theory.

Since the objective is to study dynamical systems in an analogue of the Bayesian formulation, it is natural to consider collections $\{\uvx_k\}_{k \geq 1}$ of uncertain variables with $k$ a time-like index and with $\uvx_k$ representing the state of the targets of interest in $\sfX$ at time $k$. The collection $\{\uvx_k\}_k$ can be referred to as an \emph{uncertain process}. It is also convenient to assume some form of independence between uncertain variables at different times. If the uncertain variables in the uncertain process $\{\uvx_k\}_k$ are pairwise weakly-independent and described by the same possibility function $f_{\uvx}$, then they are said to be \emph{independently identically described} (i.i.d.) by $f_{\uvx}$. We also use the abbreviation i.i.d.\ since there is no possible confusion in a given context. Alternatively, if for any $n \geq 0$ it holds that
\eqns{
f_{\uvx_k|\uvx_{1:k-1}}(x_k \given x_{1:k-1}) = f_{\uvx_k|\uvx_{k-1}}(x_k \given x_{k-1})
}
for any $x_1,\dots,x_k \in \sfX$, then $\{\uvx_k\}_k$ is said to be an \emph{uncertain Markov process}. This notion allows for defining an analogue of the concept of hidden Markov model, which, when combined with \eqref{eq:BayesRule2}, leads to an alternative formulation of single-target filtering. The additional notions introduced in the remainder of this article will allow for addressing more complex problems including uncertainty on the number of false alarms at each time step and on the presence of the target in the area of interest.

\section{Uncertain counting measure}
\label{sec:uncertainCountingMeasure}

In multi-target tracking, there is an inherent need to model that targets come in uncertain number and with uncertain states. It is therefore natural to consider the concept of point process \cite{Daley2003} (a.k.a.\ random counting measure). Formally, a point process $\calX$ is a measure-valued random variable such that $\calX(B)$ is the (random) number of points within a given set\footnote{measure-theoretic details will be omitted} $B \subseteq \sfX$. This is the reason for the alternative name ``random counting measure'': $\calX$ is indeed a measure such that $\calX(B)$ counts the number of points in $B$. For instance, $\calX(\sfX)$ is simply the total number of points in the point process $\calX$, which is also random. The most common way of writing a point process is based on Dirac measures; an example of such a Dirac measure is $\delta_x$ for a given point $x \in \sfX$, which is such that $\delta_x(B)$ equals $1$ is $x \in B$ and $0$ otherwise, for any $B \subseteq \sfX$. The point process $\calX$ can then be expressed as
$$
\calX = \sum_{i=1}^N \delta_{X_i},
$$
where $N$ is the random number of points in $\calX$ and where $X_i$ is a random variable in $\sfX$, $i \in \{1,\dots,N\}$. When assuming that the point process $\calX$ is \emph{simple}, i.e.\ $X_i \neq X_j$ almost surely for any $i \neq j$, it is also possible to identify $\calX$ with the random set $\{X_1,\dots,X_N\}$.

Point processes and the corresponding random sets have been applied to multi-target tracking over the last 20 years with undeniable success \cite{Mahler2003, Mahler2007, Stone2013}. In this context, each point in the considered point process is interpreted as the state of a target.
In order to proceed in this direction, we have to introduce an analogue of the notion of point process based on uncertain variables, which we refer to as \emph{uncertain counting measure}. Let $\calX$ be such an uncertain counting measure on $\sfX$, defined as
\eqns{
\calX = \sum_{i=1}^{\uvn} \delta_{\uvx_i}
}
where $\uvn$ is the uncertain number of points in $\calX$ and where $\uvx_i$ is an uncertain variable in $\sfX$ for any $i \in \{1,\dots,\uvn\}$. The uncertain variable $\uvn$ is described by a possibility function $f_{\uvn}$ on $\bbN_0$. Given that $\uvn = n$, the uncertain variable $(\uvx_1,\dots,\uvx_n)$ on $\sfX^n$ is described by the possibility function $f_{\calX}( \cdot \given n)$. This possibility function is assumed to be symmetrical, i.e.\
\eqns{
f_{\calX}(x_1,\dots,x_n \given n) = f_{\calX}\big( x_{\sigma(1)},\dots,x_{\sigma(n)} \Given n \big)
}
for any $n > 0$ and any permutation $\sigma$ of $\{1,\dots,n\}$. The unconditional possibility function describing $\calX$ is then defined as $f_{\calX}(x_1,\dots,x_n) = f_{\uvn}(n) f_{\calX}(x_1,\dots,x_n \given n)$ for any $n > 0$ and any $x_1,\dots,x_n \in \sfX$ and as $f_{\calX}(\psi_0) = f_{\uvn}(0)$ where $\psi_0$ is an isolated state representing the fact that there are no points in $\calX$. The possibility function $f_{\calX}$ is therefore defined on the extended set $\bar{\bbX} = \{\psi_0\} \cup \bbX$ with $\bbX = \bigcup_{n > 0} \sfX^n$. We will use a slight abuse of notations and consider that
$$
\sup_{(x_1,\dots,x_n) \in \bbX} f_{\calX}(x_1,\dots,x_n)
$$
means that the supremum is taken over all $(x_1,\dots,x_n) \in \sfX^n$ \emph{and} over all $n > 0$, and similarly for supremums over $\bar{\bbX}$.

The standard concept of independence remains relevant for uncertain counting measures. In particular, an uncertain counting measure is said to be i.i.d.\ if there exists a possibility function $f_{\uvx}$ on $\sfX$ such that
\eqns{
f_{\calX}(x_1, \dots, x_n \given n) = \prod_{i=1}^n f_{\uvx}(x_i)
}
for all $x_1,\dots,x_n$ in $\sfX$ and for any $n > 0$. However, there is no natural equivalent of the Poisson distribution and therefore the simplest form of uncertain counting measure will not be the analogue of a Poisson point process. Yet, one could for instance model the complete absence of information about the number of points $\uvn$ in the uncertain counting measure $\calX$, which would be described by $f_{\uvn}(n) = 1$ for any $n \geq 0$.

For a fixed subset $B$ of $\sfX$, one can consider the uncertain variable $\calX(B)$ on $\bbN_0$, that is $\calX(B)$ is equal to the number of points of $\calX$ in the subset $B$. The importance of this uncertain variable is linked to the concept of first-moment measure, a.k.a.\ \emph{intensity} measure, which is defined as the expectation of the corresponding quantity for point processes. There are two ways of extending this notion to uncertain counting measures: either as $\bar\bbE(\calX(B))$ or as $\bbE^*(\calX(B))$ (the former is well defined since we are dealing with integers). However, neither of these quantities have desirable properties and a further modification is required: since it holds that $\bar\bbE(\calX(B)) = \bar\bbE\big( \sum_{i=1}^{\uvn} \ind{B}(\uvx_i) \big)$, one could follow the same motivation as before and replace the sum by a maximum to obtain
$$
\bar{F}_{\calX}(B) = \bar\bbE\Big( \max_{i \in \{1,\dots, \uvn\}} \ind{B}(\uvx_i) \Big), \qquad B \subseteq \sfX.
$$
The function $\bar{F}_{\calX}$ which is defined on all subsets of $\sfX$ can be easily verified to be an outer measure. The meaning of $\bar{F}_{\calX}$ is made more apparent by the following \lcnamecref{res:outerIntensityMeasure}.

\begin{proposition}
\label{res:outerIntensityMeasure}
Let $\calX$ be an uncertain counting measure on $\sfX$, then the outer measure $\bar{F}_{\calX}$ on $\sfX$ associated with $\calX$ verifies $\bar{F}_{\calX}(B) = \bar{\bbP}(\calX(B) > 0)$ for any subset $B$ of $\sfX$.
\end{proposition}

Since the outer measure $\bar{F}_{\calX}$ evaluated at $B$ is the credibility of the fact that $\calX$ has at least one point in $B$, it follows that $\bar{F}_{\calX}(\sfX) = \bar{\bbP}(\uvn > 0)$ so that $\bar{F}_{\calX}$ is not an o.p.m.\ in general. The expression of the outer measure $\bar{F}_{\calX}$ can be given more explicitly as
\eqnsa{
\bar{F}_{\calX}(B) & = \sup_{(x_1,\dots,x_n) \in \bbX} \Big( \max_{i \in \{1,\dots, n\}} \ind{B}(x_i) f_{\calX}(x_1,\dots,x_n) \Big) \\
& = \sup_{(x_1,\dots,x_n) \in \bbX: x_1 \in B} f_{\calX}(x_1, \dots, x_n ).
}
This outer measure can also be characterised point-wise by a \emph{presence function} $F_{\calX}$ on $\sfX$ defined as
$$
F_{\calX}(x) = \sup_{(x_2,\dots,x_n) \in \bar{\bbX}} f_{\calX}(x, x_2,\dots,x_n), \qquad x \in \sfX.
$$
The supremum $\ninf{F_{\calX}}$ of the presence function is the credibility of the fact that there is at least one point in the uncertain counting measure $\calX$, indeed, it holds that $\ninf{F_{\calX}} = \max_{n > 0} f_{\uvn}(n) = \bar{\bbP}(\uvn > 0)$. In the following \lcnamecref{thm:sumIndUcm}, we give the form of the presence function of the sum of two weakly-independent uncertain counting measures.

\begin{theorem}
\label{thm:sumIndUcm}
Let $\calX$ and $\calX'$ be two weakly-independent uncertain counting measures on $\sfX$, then the presence function $F_{\calZ}$ of the uncertain counting measure $\calZ = \calX + \calX'$ is characterised by
$$
F_{\calZ}(x) = \max \{ F_{\calX}(x), F_{\calX'}(x) \}, \qquad x \in \sfX.
$$
\end{theorem}

The next theorem shows how to apply a dynamical model to a presence function, which is also a crucial step in multi-target tracking when predicting the state of targets at a given time based on the presence function at a previous time. For this purpose, we introduce another uncertain counting measure $\calZ$ on a set $\sfZ$ defined as follows: for a given realisation $\sum_{i=1}^n \delta_{x_i}$ of $\calX$, the uncertain counting measure $\calZ = \sum_{i=1}^n \delta_{\uvz_i}$ is such that
$$
\uvz_i = G(x_i) + \uvu_i, \qquad i \in \{1,\dots,n\},
$$
where the collection of uncertain variables $\{\uvu_i\}_{i=1}^n$ is i.i.d.\ by $f_{\uvu}$. The credibility of $\uvz_i = z_i$ given $\uvx_i = x_i$ is denoted $g(z_i \given x_i)$ and is equal to $f_{\uvu}(z_i - G(x_i))$.

\begin{theorem}
\label{thm:predictionFirstMoment}
The presence function $F_{\calZ}$ of the uncertain counting measure $\calZ$ verifies
\eqns{
F_{\calZ}(z) = \sup_{x \in \sfX} g(z \given x) F_{\calX}(x), \qquad z \in \sfZ.
}
\end{theorem}

\Cref{thm:sumIndUcm,thm:predictionFirstMoment} confirm that the definition of presence function for uncertain counting measures preserves some fundamental properties of the concept of first-moment measure for point processes. In particular, when combining \Cref{thm:sumIndUcm} and \Cref{thm:predictionFirstMoment}, one can obtain the predicted presence function describing multiple targets that have been propagated from the last time step to the current time step plus some newborn targets.

We now consider that the uncertain counting measure $\calX$ is i.i.d.\ by the possibility function $f_{\uvx}$, i.e.\ each point in $\calX$ is independently described by $f_{\uvx}$. The expression of $F_{\calX}$ simplifies to $F_{\calX}(x) = f_{\uvx}(x)\max_{n > 0} f_{\uvn}(n)$. Conversely, if the only information about an uncertain counting measure $\calX$ is given by a presence function $F_{\calX}$ then one can define a compatible possibility functions $f_{\uvn}$ describing the number of points in $\calX$ as
\begin{equation}
\label{eq:recoverCardinality}
f_{\uvn}(n) =
\begin{cases*}
1 & if $n = 0$ \\
\ninf{F_{\calX}} & otherwise.
\end{cases*}
\end{equation}
Indeed, the only information we obtain from $F_{\calX}$ about the number of points $\uvn$ in $\calX$ is that the credibility of having more than one point is $\ninf{F_{\calX}}$. The possibility function $f_{\uvn}$ defined in \eqref{eq:recoverCardinality} is an upper bound for all symmetrical possibility functions describing $\uvn$ which could have induced $F_{\calX}$ in the first place. If $\ninf{F_{\calX}} = 0$ then the uncertain counting measure $\calX$ does not contain any points and there is no spatial possibility function to recover so we now assume that $\ninf{F_{\calX}} > 0$. If it is known that $F_{\calX}$ was computed based on an independent description of $\calX$ then one could recover $f_{\uvx}$ as
\begin{equation}
\label{eq:recoverCommonFunction}
f_{\uvx}(x) = \dfrac{F_{\calX}(x)}{\ninf{F_{\calX}}}, \qquad x \in \sfX.
\end{equation}
In the absence of any knowledge about the type of possibility function that induced $F_{\calX}$, we cannot exclude correlations and have to define $f_{\calX}$ as 
\begin{equation}
\label{eq:recoverFunction}
f_{\calX}(x_1,\dots,x_n) = \big( F_{\calX}(x_1) \dots F_{\calX}(x_n) \big)^{1/n},
\end{equation}
for any $(x_1,\dots,x_n) \in \sfX^n$ and any $n > 0$. Once again, $f_{\calX}$ is an upper bound for all possibility functions describing $\calX$ which could have induced $F_{\calX}$.

\begin{example}
\label{ex:maxCorrelated}
Let $\uvx_1,\dots,\uvx_n$ be a collection of uncertain variables on $\sfX$ and let $\uvz$ an uncertain variable on $\sfZ$ such that $\uvx_i = T_i(\uvz)$ for a given map $T_i : \sfZ \to \sfX$, $i \in \{1,\dots,n\}$. If $f_{\uvz}$ describes $\uvz$ then it follows from the change of variable formula \eqref{eq:changeOfVariable} that the possibility function $f_{\uvx_{1:n}}$ describing $(\uvx_1,\dots,\uvx_n)$ is characterised by
$$
f_{\uvx_{1:n}}(x_{1:n}) = \sup_{z \in \sfZ} f_{\uvz}(z) \prod_{i=1}^n \ind{T_i(z)}(x_i), \qquad x_{1:n} \in \sfX^n.
$$
This is an example where $\uvx_1,\dots,\uvx_n$ are maximally correlated. We denote by $f_{\uvx_i}$ the marginal possibility function $\uvx_i$ which verifies
$$
f_{\uvx_i}(x_i) = \sup_{z \in \sfZ} f_{\uvz}(z) \ind{T_i(z)}(x_i) = \sup_{z \in \sfZ: T_i(z) = x_i} f_{\uvz}(z),
$$
for any $x_i \in \sfX$. Fixing $z \in \sfZ$ and setting $x_i = T_i(z)$ for any $i \in \{1,\dots,n\}$ leads to
$$
\prod_{i=1}^n f_{\uvx_i}(x_i) = f_{\uvz}(z)^n = f_{\uvx_{1:n}}(x_{1:n})^n.
$$
Therefore, when recovering $f_{\uvx_{1:n}}$ from its marginals, we take roots such as $f_{\uvx_{1:n}}(x_{1:n}) = f_{\uvx_1}(x_1)^{\omega_1} \dots f_{\uvx_n}(x_n)^{\omega_n}$, with $\omega_1 + \dots + \omega_n = 1$ to avoid redundancy. This is consistent with \eqref{eq:recoverFunction} when $\omega_1 = \dots = \omega_n = 1/n$, as required by the assumption of symmetry.
\end{example}

Since \Cref{ex:maxCorrelated} studies maximally-correlated uncertain variables, it follows that
\begin{equation}
\label{eq:possibilityFromMarginals}
f_{\uvx_{1:n}}(x_{1:n}) \leq f_{\uvx_1}(x_1)^{\omega_1} \dots f_{\uvx_n}(x_n)^{\omega_n}
\end{equation}
in general. This upper bound must therefore be considered when there is no additional information about correlations between $\uvx_i$ and $\uvx_j$, $i \neq j$. It follows from the definition \eqref{eq:variance} of the variance $\bbV^*(\bscdot)$ that taking the $r$th root of a possibility function multiplies the corresponding variance by $r$. Thus, another interpretation of \eqref{eq:possibilityFromMarginals} is that (weak) independence can always be obtained by loosing a sufficient amount of information. Taking the root of a possibility function is also justified by the fact that the resulting function is still a possibility function; this is not the case with probability distributions.

It follows from \eqref{eq:recoverCardinality} and \eqref{eq:recoverCommonFunction} that the only possibility functions describing $\calX$ which can be recovered exactly from the presence function $F_{\calX}$ are the ones that describe points of $\calX$ independently and for which $f_{\uvn}(0) = 1$ and $f_{\uvn}$ is constant over the set $\bbN$ of positive integers. Henceforth, only independently described uncertain counting measures will be considered so that \eqref{eq:recoverCommonFunction} will always be used.

Being equipped with a way of describing imprecise information about multiple targets, we proceed to the modelling of their dynamics and observation in order to enable the derivation of filtering equations.

\section{Model}
\label{sec:model}

We aim to define the analogue of the standard multi-target model in the context of possibility theory. A more sophisticated model relying on both uncertain and random variables could be defined in situations where some aspects of the model can be faithfully described through probability theory; we however focus on the fully-possibilistic case for the sake of simplicity. Without loss of generality, time is assumed to take integer values. In order to improve readability, we will write $a \lor b$ instead of $\max\{a,b\}$ for any $a,b \in \bbR$. We also consider that $\lor$ has a lower precedence than multiplication so that $ a\lor bc = a \lor (bc)$ for any $a,b,c \in \bbR$.

\subsection{Dynamics}

Let the state space $\sfX$ be the union of a subset $S$ of $\bbR^d$, for some $d > 0$, with an isolated state $\psi$ representing the case where the target does not admit a state in $S$. The state $\psi$ relates to the \emph{absence} of the target from $S$, e.g.\ because its position is out of the bounds of the considered area; it allows for modelling the birth/death of a target as a simple change of state between any point of $S$ and $\psi$. At any given time $k \in \bbN$, the uncertainty in the state of a target is modelled by an uncertain variable $\uvx_k$ on $\sfX$. It is assumed that the collection $\{\uvx_k\}_k$ is an uncertain Markov process so that the corresponding state equation on $S$ can be expressed as
\eqns{
\uvx_k = G(\uvx_{k-1}) + \uvu_k
}
where $G : S \to S$ is a given map related to the dynamics and where the collection of uncertain variables $\{\uvu_k\}_k$ on $S$ is i.i.d.\ by the possibility function $f_{\uvu}$. The possibility function describing $\uvx_k$ is denoted by $g_k(\cdot \given \uvx_{k-1})$ and is characterised on $S \times S$ by $g_k(x_k \given x_{k-1}) = f_{\uvu}(x_k - G(x_{k-1}))$ for all $x_k, x_{k-1} \in S$. The scalar $g_k(\psi \given x_{k-1})$ is the credibility for a target at $x_{k-1} \in S$ not to survive from one time step to the other while $g_k(x_k \given \psi)$ is the credibility for a target to be born at $x_k \in S$. The dependence of $g_k(\cdot \given \uvx_{k-1})$ on the time step $k$ is assumed to be only through $g_k(\cdot \given \psi)$. Finally, $g_k(\psi \given \psi)$ is assumed to be equal to $1$ and can be interpreted as the credibility for a target to remain ``unborn''.

As opposed to the corresponding probabilistic modelling, the transition function $g_k(\bscdot \given x_{k-1})$ is not supposed to be a full characterisation of the dynamics; instead, this function is interpreted as a description of the most extreme dynamics that are expected to be observed. This interpretation is motivated by the fact that $g_k(\bscdot \given x_{k-1})$ can be seen as an upper bound for subjective transition p.d.f.s $p_k(\bscdot \given x_{k-1})$ in the sense that
\begin{equation}
\label{eq:upperBoundTransition}
\int_B p_k(x_k \given x_{k-1}) \d x_k \leq \sup_{x_k \in B} g_k(x_k \given x_{k-1}), \qquad B \subseteq S,
\end{equation}
for any $x_{k-1} \in S$. This interpretation is standard in possibility theory and Dempster-Shafer theory. Importantly, there is not necessarily a true p.d.f.\ $p_k(\bscdot \given x_{k-1})$ and the inequality \eqref{eq:upperBoundTransition} only serves as a way to interpret and define $g_k(\bscdot \given x_{k-1})$. In fact, this worst-case approach is often the one that is considered by target-tracking practitioners when defining a model and the proposed framework appears to be consistent with this.

Newborn targets at time $k$ are modelled by an uncertain counting measure $\calX_{\b,k}$, weakly-independent of all other quantities. The corresponding presence function is denoted by $F_{\b,k}(x_k) = g_k(x_k \given \psi)$, $x_k \in \sfX$. The presence function $F_{\b,k}$ is in fact a possibility function since $g_k(\bscdot \given \psi)$ has supremum $1$ by construction (it is itself a possibility function). The uncertain counting measure on $\sfX$ made of the states of all targets at time $k$ is denoted $\calX_k$.

\subsection{Observation}
\label{sec:observation}

Let the observation space $\sfY$ be the union of a subset $S'$ of $\bbR^{d'}$, for some $d' > 0$, with an isolated point $\phi$ representing the case where a target does not produce any observation. The point $\phi$, which can be interpreted as an empty observation, allows for modelling detection failures in the same way as actual detection: the event ``the target has not been detected'' becomes ``the target has generated the observation $\phi$''. The considered definition of the observation space $\sfY$ allows for modelling the observation of a given target as a standard filtering problem instead of introducing point processes (or the corresponding random sets) with either $0$ or $1$ point as is usual \cite{Mahler2007}; this approach was also used in \cite{Singh2009,Caron2011}. The observation of a given target is represented by an uncertain variable $\uvy_k$ in $\sfY$. The observation $\uvy_k$ is assumed to be conditionally independent of all other observations given the state $\uvx_k$ of the considered target at time $k$. It follows that the observation equation can be expressed as
\eqnl{eq:observationEquation}{
\uvy_k = H(\uvx_k) + \uvv_k
}
for a given map $H : S \to S'$, where the collection of uncertain variables $\{\uvv_k\}_k$ on $S'$ is i.i.d.\ by the possibility function $f_{\uvv}$. The possibility function describing $\uvy_k$ is the likelihood function denoted by $h(\cdot \given \uvx_k)$, which is characterised on $S' \times S$ by $h(y_k \given x_k) = f_{\uvv}(y_k - H(x_k))$ for all $y_k \in S'$ and all $x_k \in S$. The scalar $h(\phi \given x_k)$ is the credibility for the detection of a target at $x_k \in S$ to fail and $h(y_k \given \psi)$ is assumed to be equal to $0$ if $y_k \in S'$ and to $1$ if $y_k  = \phi$; indeed, targets that are not in the state space cannot be detected. With this model, the analogue of the probability of detection is the possibility of detection at $x \in \sfX$ which is denoted $\alpha_{\d}(x)$ and defined as
$$
\alpha_{\d}(x) = \sup_{y \in S'} h(y \given x).
$$
Conversely, the possibility of detection failure at $x \in \sfX$ is defined as $\alpha_{\df}(x) = h(\phi \given x)$. As opposed to the probabilistic context, $\alpha_{\df}(x)$ cannot be deduced from $\alpha_{\d}(x)$ since these quantities verify $\alpha_{\d}(x) \lor \alpha_{\df}(x) = 1$ rather than $\alpha_{\d}(x) + \alpha_{\df}(x) = 1$. The functions $\alpha_{\d}$ and $\alpha_{\df}$ induce upper and lower bounds for the probability of detection $p_{\d}$ as
$$
1 - \alpha_{\df}(x) \leq p_{\d}(x) \leq \alpha_{\d}(x), \qquad x \in \sfX.
$$
These bounds facilitate the interpretation of $\alpha_{\d}$ and $\alpha_{\df}$ as follows: if we believe that the probability of detection at $x \in \sfX$ should be greater than, say $0.7$, then we can set $\alpha_{\d} = 1$ and $\alpha_{\df} = 0.3$. This also shows how the considered framework brings modelling flexibility. The same analysis can be applied to the transition function via the corresponding credibility of (non)survival. 

False alarms at time $k$ are modelled by an uncertain counting measure $\calY_{\fa,k}$, weakly-independent of all other quantities. The corresponding presence function on $\sfY$ is assumed to be time-invariant and is therefore denoted $F_{\fa}$; we assume that $F_{\fa}(\phi) = 1$ for convenience since this turns $F_{\fa}$ into a possibility function. We denote by $\bar{\calY}_k$ the uncertain counting measure including all observations at time $k$, i.e.\
$$
\bar{\calY}_k = \calY_{\fa,k} + \sum_{i=1}^{\uvm_k} \delta_{\uvy_{i,k}}
$$
where $\uvm_k$ is the uncertain number of targets and where $\uvy_{i,k}$ is the observation generated by the $i$th target in $\sfY$ (therefore including the point $\phi$). We do not actually observe the entirety of $\bar{\calY}_k$ but rather its restriction to $S'$, which we denote $\calY_k$. The realisation of $\calY_k$ is interpreted as a set $Y_k = \{y_{1,k}, \dots, y_{m_k,k}\}$ for some given $m_k \in \bbN_0$.

\begin{remark}
Considering $F_{\fa}(\phi) = 1$ is natural when seeing false alarms as observation generated by objects that are not targets (sometimes referred to as \emph{false-alarm generators}). With this modelling, it is clear that potentially many of these undesired objects will actually fail to generate any actual observation (which would be a false alarm for us) so their observation is indeed $\phi$.
\end{remark}

The defined single-target models for dynamics and observation naturally lead to the introduction of an analogue of the Kalman filter in the context of possibility theory as derived in \cite{Houssineau2018}. Remarkably, this alternative Kalman filter displays the same predicted/posterior expected values and variances as the original. There are however differences between the two formulations, as explained in the following section. Related but different results can be found in \cite{Oussalah2000} and in \cite{Mahler2007}. A non-linear single-target filtering problem is also considered in \cite{Ristic2018}, which relies on an approximation of possibility function based on Monte Carlo methods \cite{Houssineau2017}. This is extended to tracking a single-target with detection failures and false alarms in \cite{Ristic2019}, where \emph{uncertain finite sets} are used instead of uncertain counting measures. Note that presence functions are not needed when tracking a single target.


\subsection{Spatial necessity}
\label{sec:spatial_necessity}

If we consider a prior possibility function $f_{\uvx_k}$ on $S$ and an observation $y_k \in S'$ at time $k$ then the corresponding marginal likelihood $f_{\uvy_k}(y_k) = \sup_{x \in S} h(y_k \given x) f_{\uvx_k}(x)$ indicates the level of agreement between the prior $f_{\uvx_k}$ and the observation $y_k$ via the likelihood $h$ as opposed to its probabilistic counterpart which indicates the \emph{fitness} of the prior. The best example of this difference is when the prior is uninformative in which case the probabilistic marginal likelihood will usually be very small whereas $f_{\uvy_k}(y_k)$ will be equal to $1$.

In some cases, a notion of fitness can be useful when working with possibility functions. To obtain such a quantity, one can introduce a subjective probability $p$ on $S$ and use a double inequality of the form
$$
1 - \sup_{x \in S} (1-\varphi(x)) f_{\uvx_k}(x) \leq \int \varphi(x) p(x) \d x \leq \sup_{x \in S} \varphi(x) f_{\uvx_k}(x)
$$
which holds for any integrable function $\varphi$ on $S$ and which is a generalisation of \eqref{eq:upperBoundTransition}. In particular, if $\varphi(x) = h(y_k \given x)$ then the upper bound is the marginal likelihood $f_{\uvy_k}(y_k)$ and the lower bound can be interpreted as the necessity for $y_k$ to originate from a target described by $f_{\uvx_k}$. This quantity behaves according to intuition, i.e.\ if $f_{\uvx_k}(x)$ is equal to $0$ everywhere except at the maximum likelihood $\argmax_{x \in \sfX} h(y_k \given x)$ then the necessity is equal to $1$ whereas if $f_{\uvx_k}$ is uninformative then the necessity is equal to $0$. The gap between the necessity and the possibility quantifies an additional level of uncertainty which can be interpreted as an interval for the probabilistic marginal likelihood when the subjective prior p.d.f.\ $p$ is upper bounded by $f_{\uvx_k}$. In other words, this gap characterises the effect of the uncertainty represented by $f_{\uvx_k}$ on the fitness with respect to the likelihood $h$, i.e.\ we assume that the likelihood $h$ is correct and test the prior $f_{\uvx_k}$. If, for instance, we wanted to test the fitness of the prior $f_{\uvx_{k-1}}$ against the observation $y_k$ then we would need to consider $\varphi(x) = \sup_{x' \in S} h(y_k \given x') g_k(x' \given x)$. Computing the necessity will be useful when performing track extraction as detailed in \Cref{sec:track_extraction}.

\section{Recursion}
\label{sec:recursion}

Having introduced a seemingly-appropriate analogue of intensity function as well as a model of targets' dynamics and observation, we now aim to derive a recursion for the predicted and posterior presence function of $\calX_k$. We denote by $F_{k-1}(\bscdot \given Y_{1:k-1})$ the posterior presence function representing the uncertain counting measure $\calX_{k-1}$ on $\sfX$ given the observation up to time $k-1$. We then assume that $F_{k-1}(\bscdot \given Y_{1:k-1})$ is available and seek to express the posterior presence function $F_k(\bscdot \given Y_{1:k})$ as a function of it. There is no particular reason to believe that the latter is sufficient to compute the former without any other information about the possibility function on $\bar{\bbX}$ describing $\calX_{k-1}$; yet, we show in this section that this is indeed the case.

We now consider the computation of predicted presence functions. The following \lcnamecref{cor:predictedIntensity}, which specifies how to predict the presence function from time $k-1$ to time $k$, is a direct consequence of Theorems~\ref{thm:sumIndUcm} and \ref{thm:predictionFirstMoment}.

\begin{corollary}
\label{cor:predictedIntensity}
The predicted presence function $F_k(\bscdot \given Y_{1:k-1})$ is characterised by
\begin{equation}
\label{eq:predictedIntensity}
F_k(x_k \given Y_{1:k-1}) = F_{\b,k}(x_k) \lor \sup_{x \in S} g_k(x_k \given x) F_{k-1}(x \given Y_{1:k-1})
\end{equation}
for any $x_k \in \sfX$.
\end{corollary}

Based on the definition of $F_{\b,k}$, it appears that $F_k(\psi \given Y_{1:k-1})$ is equal to $1$, so that this presence function is actually a possibility function. Since there is usually no objection against target survival, we will assume that the credibility $\alpha_{\s}(x) = \sup_{x_k \in S} g_k(x_k \given x)$ for a target at $x \in S$ to survive to the next time step is equal to $1$. Conversely, the credibility $\alpha_{\ns}(x) = g_k(\psi \given x)$ for a target at $x \in S$ not to survive to the next time step is usually small and it is natural to assume that $\alpha_{\ns}(x) \ll 1$. However, the function $\alpha_{\ns}$ has no bearing on the prediction equation \eqref{eq:predictedIntensity} and does not actually need to be defined. Indeed, as discussed previously with $\alpha_{\d}$ and $\alpha_{\df}$, the functions $\alpha_{\s}$ and $\alpha_{\ns}$ are not directly related and must simply verify $\alpha_{\s}(x) \lor \alpha_{\ns}(x) = 1$.

The form of \eqref{eq:predictedIntensity} informs us about the necessary amount of information regarding newborn targets. For instance, setting $F_{\b,k}(x_k) = 1$ for all $x_k \in S$, which would model the total absence of information about newborn targets, would imply that $F_k(\bscdot \given Y_{1:k-1})$ does not retain any of the information from the previous time step and also becomes completely uninformative. Since this is clearly to be avoided with the current recursion, we conclude that $F_{\b,k}$ should take values that are much smaller than $1$ for the prediction to hold any information. Successfully using an uninformative form for $F_{\b,k}$ would require the propagation of separate information about specific tracks. However, one can consider that $F_{\b,1}(x) = 1$ for all $x \in \sfX$ in order to model that an arbitrary number of targets might already be present when turning on the sensor.

We now consider the problem of updating presence functions in the following \lcnamecref{thm:updatedIntensity}.

\begin{theorem}
\label{thm:updatedIntensity}
The posterior presence function $F_k(\bscdot \given Y_{1:k})$ is characterised by
\begin{multline}
\label{eq:updatedIntensity}
F_k(x_k \given Y_{1:k}) = h(\phi \given x_k) F_k(x_k \given Y_{1:k-1}) \\
\lor \max_{y \in Y_k} \dfrac{h(y \given x_k)F_k(x_k \given Y_{1:k-1})}{F_{\fa}(y) \lor \sup_{x \in \sfX} h(y \given x)F_k(x \given Y_{1:k-1})}
\end{multline}
for any $x_k \in \sfX$.
\end{theorem}

The proof of \Cref{thm:updatedIntensity}, which can be found in Appendix~\ref{proof:updatedIntensity}, is inspired from \cite{Singh2009,Caron2011}, see also \cite{DelMoral2015}. Since it holds that $h(\phi \given \psi) = 1$ and that $F_k(\psi \given Y_{1:k-1}) = 1$, it follows that the posterior presence function also verifies $F_k(\psi \given Y_{1:k}) = 1$. Therefore, all the considered presence functions are actually possibility functions.

Equation~\eqref{eq:updatedIntensity} informs us on a way to define the presence function $F_{\b,k}$; indeed, for a given $y \in Y_k$, the scalar $e_{\b,k} = \sup_{x \in S} h(y \given x) F_{\b,k}(x) \in [0,1]$ is the possibility for $y$ to be the first observation of a target. In a simple setting where $F_{\b,k}$ is constant on $S$ and equal to $\alpha_{\b}$ and where $\sup_{x \in S} h(y \given x) = 1$, the expression of $e_{\b,k}$ simplifies to $e_{\b,k} = \alpha_{\b}$.

The recursion defined by \Cref{cor:predictedIntensity} and \Cref{thm:updatedIntensity} is naturally reminiscent of the PHD filter \cite{Mahler2003}. The latter however assumes that the point process to be inferred as well as the false alarms are Poisson i.i.d., whereas the proposed recursion makes no assumption about the parametric form of the underlying uncertain counting measures. Yet, both approaches require some form of approximation. One main difference is that the posterior presence function provides no specific information about the number of targets; indeed, $\hat{e}_k = \sup_{x_k \in S} F_k(x_k \given Y_{1:k})$ is simply the credibility that there is at least one target.
The proposed approach must therefore rely on track extraction to estimate the number of targets.

\begin{remark}
\label{rem:extension}
A connection between the proposed recursion and the standard Bayesian filtering equations can be made by extending $\sfX$ to $\bar{\sfX} = \sfX \cup \{\psi_{\fa}\}$ with $\psi_{\fa}$ another isolated state representing false-alarm generators. Then, recalling that $F_{\b,k}(x_k) = g_k(x_k \given \psi)$, $x_k \in \sfX$, and defining $\bar{Y}_k$ as the extended set of observations $Y_k \cup \{\phi\}$, the expressions of the predicted and posterior possibility functions can then be simplified to
\begin{align*}
F_k(x_k \given Y_{1:k-1}) & = \sup_{x \in \bar{\sfX}} g_k(x_k \given x) F_{k-1}(x \given Y_{1:k-1}) \\
F_k(x_k \given Y_{1:k})  & = \max_{y \in \bar{Y}_k} \dfrac{h(y \given x_k)F_k(x_k \given Y_{1:k-1})}{\sup_{x \in \bar{\sfX}} h(y \given x)F_k(x \given Y_{1:k-1})}
\end{align*}
for any $x_k \in \bar{\sfX}$, when extending the different terms appropriately, i.e.\ $F_{k-1}(\psi_{\fa} \given Y_{1:k-1}) = 1$, $g_k(\psi_{\fa} \given x) = g_k(x \given \psi_{\fa}) = 0$ for any $x \in \sfX$ and $h(y \given \psi_{\fa}) = F_{\fa}(y)$. Although these compact expressions are useful in proofs, the more explicit expressions \eqref{eq:predictedIntensity} and \eqref{eq:updatedIntensity} are preferred in general.
\end{remark}

The recursion \eqref{eq:predictedIntensity} and \eqref{eq:updatedIntensity} can be simplified by considering the following assumptions:
\begin{enumerate}[label=(\roman*)]
\item \label{it:constantBirth} the presence function $F_{\b,k}$ is constant over $S$, i.e.\ $F_{\b,k}(x) = \alpha_{\b}$ for any $x \in S$
\item \label{it:constantDetectionFailure} the credibility of a detection failure does not depend on the state, i.e.\ $h(\phi \given x) = \alpha_{\df} < 1$ for any $x \in S$, and verifies $\alpha_{\df} < 1$.
\item \label{it:constantFalseAlarm} the presence function $F_{\fa}$ is also constant on $S'$, i.e.\ $F_{\fa}(y) = \alpha_{\fa}$ for any $y \in S'$
\end{enumerate}
Under these assumptions, the filtering equations \eqref{eq:predictedIntensity} and \eqref{eq:updatedIntensity} become
\begin{align*}
F_k(x_k \given Y_{1:k-1}) & = \alpha_{\b} \lor \sup_{x \in \sfX} g_k(x_k \given x) F_{k-1}(x \given Y_{1:k-1}) \\
F_k(x_k \given Y_{1:k}) & = \alpha_{\df} F_k(x_k \given Y_{1:k-1}) \\
& \quad \lor \max_{y \in Y_k} \dfrac{h(y \given x_k)F_k(x_k \given Y_{1:k-1})}{\alpha_{\fa} \lor \sup_{x \in \sfX} h(y \given x)F_k(x \given Y_{1:k-1})}
\end{align*}
The second equation could be further simplified by assuming that there is no information at all about the false alarms, that is $\alpha_{\fa} = 1$, however that would cause $\hat{e}_k$ to inexorably decrease in time, which is not desired. We study some properties of the proposed method in the following sections.

\subsection{Behaviour}

We first highlight a few practical aspects of the recursion defined by \eqref{eq:predictedIntensity} and \eqref{eq:updatedIntensity} and of the corresponding algorithm.

\paragraph{Presence function} If we consider the case where there is considerable uncertainty on the origin of a given observation $y \in Y_k$; in particular if we assume that the marginal likelihood $\sup_{x \in \sfX} h(y \given x) F_k(x | Y_{1:k-1})$ is only slightly greater than the possibility of false alarm $F_{\fa}(y)$, then the possibility that $y$ originates from a target, which is equal to $\sup_{x \in \sfX} \tilde{F}_k(x \given y, Y_{1:k-1})$ with
\begin{equation}
\label{eq:possibilityAssociation}
\tilde{F}_k(x \given y, Y_{1:k-1}) = \dfrac{h(y \given x)F_k(x \given Y_{1:k-1})}{F_{\fa}(y) \lor \sup_{x \in \sfX} h(y \given x)F_k(x \given Y_{1:k-1})},
\end{equation}
is equal to $1$. In the standard setting, the probability that $y$ originates from a target would only be slightly greater than $0.5$. This fact highlights the differences between the typical values taken by the presence function and the intensity function.

\paragraph{Nearby observations} Due to the form of \eqref{eq:updatedIntensity} with a maximum over observations, it follows that if two observations are arbitrarily close to each other then the proposed method gives essentially the same result as if only one of the two observations was present. This is consistent with the fact that the proposed recursion does not attempt to estimate the number of targets and only focuses on the presence of \emph{at least} one target at a given point of $S$. In the PHD filter, if the predicted number of targets is $1$ and if the intensity of the false-alarm point process is small, then, updating with two nearby observations will induce a posterior number of target that is close to $2$. Instead, it is simply acknowledged in the proposed approach that no cardinality estimates can be obtained directly from a presence function.

\paragraph{Gating} If we consider a given state $x_k \in S$, it follows from \eqref{eq:updatedIntensity} that only the closest observations will have an impact on $F_k(x_k \given Y_{1:k})$. It should therefore be possible to devise a \emph{gating} procedure without introducing additional errors (especially in Monte-Carlo-based implementations). 

\paragraph{Sensor ordering} The proposed recursion inherits the shortcomings of the PHD filter in terms of sensor ordering \cite{Nagappa2011}. This is due to the loss of information that occurs in the computation of the posterior presence function. Closed-form recursions will be needed in order to bypass this drawback.

\paragraph{Regional uncertainty in target number} Because of the Poisson assumption, the variance in target number in the PHD filter is equal to the mean. It is therefore necessary to set a large number of expected targets to obtain a large variance, which might yield many false tracks if a sensor starts surveying this area. Although this shortcoming can be addressed in several ways \cite{Delande2014, Schlangen2017, deMelo2018}, it remains that the original PHD filter does not suitably represent the uncertainty in regions of the state space where there is little to no information about target numbers. In particular, it is difficult to capture the uncertainty in regions that are far from the field of view of the sensor(s) and the intensity is usually set to zero there. With the suggested initialisation, the proposed approach allows for representing the fact that there could still be any number of targets outside of the field of view of the sensor(s) at any time $k \geq 1$. This flexibility is crucial in applications involving multiple sensors or a moving sensor as in \Cref{sec:movingSensor}.

\subsection{Implementation}

The recursion \eqref{eq:predictedIntensity}-\eqref{eq:updatedIntensity} can be implemented in closed form when assuming that the dynamics and observation are linear and Gaussian and that the considered presence functions are Gaussian \emph{max-mixtures} on $S$, e.g.\
\begin{equation}
\label{eq:mixture}
F_k(x \given Y_{1:k}) = \max_{i \in \{1,\dots,n_k\}} \hat{w}_{i,k} \oN(x; \hat{\mu}_{i,k}, \hat{P}_{i,k}),
\end{equation}
for any $x \in S$, where $n_k \in \bbN_0$, where $\hat{w}_{i,k} \in [0,1]$ and where $\oN(\hat{\mu}_{i,k}, \hat{P}_{i,k})$ is the $d$-dimensional generalisation of \eqref{eq:Gaussian}, with $\hat{\mu}_{i,k} \in S$ and $\hat{P}_{i,k}$ a $d \times d$ symmetric and positive-definite matrix, for any $i \in \{1,\dots,n_k\}$. As opposed to standard Gaussian mixtures, the set $S$ can be a strict subset of $\bbR^d$. The predicted expected value $\mu_{i,k}$ and variance $P_{i,k}$ as well as the posterior expected value $\hat{\mu}_{i,k}$ and variance $\hat{P}_{i,k}$ take the same expressions as in the probabilistic case \cite{Vo2006}. The recursion for the predicted and posterior weights $w_{i,k+1}$ and $\hat{w}_{i,k+1}$ does differ; the corresponding expressions follow directly from \eqref{eq:predictedIntensity} and \eqref{eq:updatedIntensity}. A pseudo-code for the proposed approach is given in Appendix~\ref{sec:pseudoCode}. Although there is no explicit data association in the proposed approach, one can record which observations have been used for updating which terms in the implementation of the algorithm as is usual with the PHD filter \cite{Ristic2010}; this will be useful for the track extraction method detailed in Section~\ref{sec:track_extraction}.

The two standard Gaussian-mixture reduction techniques \cite{Salmond1990} are pruning and merging. Whereas pruning is directly applicable to max-mixtures, merging should be applied more cautiously since max-mixtures do not behave in the same way as sum-based mixtures. In particular, only terms with similar expected value and variance can be safely merged. Yet, one advantage of Gaussian max-mixtures is that the $i$th term can be removed from the mixture without inducing any error if
$$
\hat{w}_{i,k} \oN(x; \hat{\mu}_{i,k}, \hat{P}_{i,k}) \leq \hat{w}_{j,k} \oN(x; \hat{\mu}_{j,k}, \hat{P}_{j,k}),
$$
for all $x \in S$ and for all $j \in \{1,\dots,n_k\}$. The identification of these terms is however non-trivial and is therefore out of the scope of this work.

\subsection{Track extraction}
\label{sec:track_extraction}

\subsubsection{Necessity of target presence}
\label{sec:track_extraction_1}

Classification and decision are well-known strengths of Dempster-Shafer theory \cite{Denoeux2000,Powell2006,Beynon2000}. This is due to the ability of this framework to assess the credibility of events in a more nuanced way than the standard probabilistic approach. More specifically, instead of simply providing the (subjective) probability of a given event, upper and lower bounds are provided, hence giving the choice of either immediately making a decision or waiting for more data to be collected. Possibility functions also provide this ability, albeit in a more simplistic form. This aspect can be useful in target tracking where the possible presence of a target can be signalled as soon as the upper bound reaches $1$ and then confirmed once the lower bound is above a given threshold. However, this type of operation is not directly applicable to the proposed recursion and would require the actual propagation of tracks. 
Nevertheless, it is possible to use a slightly different method for track extraction with the proposed approach: one can compute the possibility that a given observation $y$ at time $k$ originates from a false alarm as
$$
F_{\fa}(y \given Y_{1:k-1}) = \dfrac{F_{\fa}(y)}{F_{\fa}(y) \lor \sup_{x \in \sfX} h(y \given x)F_k(x \given Y_{1:k-1})},
$$
and deduce the necessity that $y$ is target-originated as $1 - F_{\fa}(y \given Y_{1:k-1})$. If this necessity is above a given threshold $\tau$, then a track can be declared with the mode of the presence function $\tilde{F}_k(\cdot \given y, Y_{1:k-1})$, defined in \eqref{eq:possibilityAssociation}, as a state. This track extraction method is sufficient in standard tracking problems, see e.g.\ Section~\ref{sec:standardScenario}; yet, additional tests might need to be carried out in general, as discussed in the following section.

\subsubsection{Spatial necessity in track extraction}
\label{sec:track_extraction_2}

When significant gaps in detection are likely, there might be many associations with large credibilities due to the large covariance of some of the terms in the predicted Gaussian max-mixture at some given time step $k$. In order to verify that these associations are not only possible but, to some extent, necessary, we assess the fitness of predicted terms against the current observations as detailed in Section~\ref{sec:spatial_necessity}. In particular, we consider a term in the Gaussian max-mixture with observations $y_{k_1},\dots,y_{k_m}$ at times $k_1,\dots,k_m$ with $k_1 < \dots < k_m < k$ and denote by $\Lambda_{k|k-1}(y ; y_{k_{1:m}})$ the necessity of the predicted possibility function $f_{\uvx_k}(\cdot \given y_{k_{1:m}})$ against the likelihood $h(y \given \cdot)$ for a given observation $y$, i.e.
$$
\Lambda_{k|k-1}(y ; y_{k_{1:m}}) = 1 - \sup_{x \in \sfX}(1 - h(y \given x)) f_{\uvx_k}(x \given y_{k_{1:m}}).
$$
The corresponding upper bound is denoted by $\Gamma_k(y_k ; y_{k_{1:m}})$. Then, before confirming this Gaussian component as a track, we can check that the gap between possibility and necessity is small enough, i.e.\ that
$$
\dfrac{\Lambda_{k|k-1}(y ; y_{k_{1:m}})}{\Gamma_k(y; y_{k_{1:m}})} > \tau_{\p},
$$
for some fixed threshold $\tau_{\p} \in (0,1)$. A similar approach can be used to assess the necessity w.r.t.\ to the initial possibility function, as described in Appendix~\ref{sec:init_spatialnecessity}, and we denote by $\tau_{\b}$ the corresponding threshold. The computational aspects associated with spatial necessities are also discussed in the appendix.

\subsection{Observation-driven birth}

One advantage of the proposed approach is that birth can be made observation-driven from a spatial viewpoint in a simple way. Indeed, the difficulty with observation-driven birth schemes in a probabilistic context is that the distribution of newborn targets cannot be made uninformative when the state space is unbounded. The probability that a given observation originates from a newborn target must then be set manually \cite{Houssineau2010, Ristic2012}.

We consider a linear-Gaussian model for the sake of simplicity, with any state $x \in S$ of the form $[\mathbf{x}, \dot{\mathbf{x}}, \mathbf{y}, \dot{\mathbf{y}}]^{\tr}$, where $[\mathbf{x}, \mathbf{y}]^{\tr}$ is the position in the 2-dimensional plane and $[\dot{\mathbf{x}}, \dot{\mathbf{y}}]^{\tr}$ is the velocity. In this situation, one can set
\begin{equation}
\label{eq:partialGaussian}
F_{\b,k}(x) = \oN\big( [\dot{\mathbf{x}}, \dot{\mathbf{y}}]^{\tr}; \dot{\mu}_{\b}, \dot{P}_{\b} \big),
\end{equation}
with $\dot{\mu}_{\b}$ and $\dot{P}_{\b}$ the expected value and variance for the velocity, respectively. The presence function $F_{\b,k}$ does not carry any information regarding the position (which will be observed) but specifies some prior knowledge about the velocity (which is hidden) as required.

\begin{remark} Equation \eqref{eq:partialGaussian} can be seen as a proper Gaussian possibility function on $S$ when parametrising it by a block-diagonal precision matrix with the zero matrix for the position and $\dot{P}_{\b}^{-1}$ for the velocity. This model is only accessible as a limit for Gaussian probability distributions \cite{Berger1992}.
\end{remark}

\section{Simulations}
\label{sec:simulations}


We simulate $T$ time steps of duration $\Delta = 1$ and implicitly assume that all units are the ones of the international system.  We consider targets evolving according to a nearly-constant-velocity model in the 2-dimensional Euclidean plane, i.e.\ $X_k = GX_{k-1} + U_k$ with $U_k \sim \N(0, Q)$ independently for any $k$, where
$$
G = I_2 \otimes
\begin{bmatrix}
1 & \Delta \\
0 & 1
\end{bmatrix}
\quad\text{and}\quad
Q = \sigma^2 I_2 \otimes 
\begin{bmatrix}
\Delta^4/4 & \Delta^3/2 \\
\Delta^3/2 & \Delta^2
\end{bmatrix},
$$
with $I_2$ the identity matrix of dimension 2 and $\otimes$ the Kronecker product. It follows that $S = \bbR^4$ and we assume that the observation space is $S' = [0, 1000] \times [0, 1000]$. The probability of survival is assumed to be state-independent and is denoted by $p_{\s}$. The position of newborn targets is uniformly distributed over the subset $S_{\mathrm{obs}} = \{x \in S : Hx \in S'\}$ of $S$ and their velocity is sampled from $\N(0,\sigma_{\b}^2 I_2)$. The number of newborn targets is Poisson distributed with parameter $\lambda_{\b} = 0.25$. The location of each target is observed with additive noise, i.e.\ $Y_k = H(X_k - x_{\s,k}) + V_k$ with
$$
H = 
\begin{bmatrix}
1 & 0 & 0 & 0 \\
0 & 0 & 1 & 0
\end{bmatrix},
$$
with $V_k \sim \N(0, \sigma'^2 I_2)$ independently for any $k$ and with $x_{s,k} \in S$ the state of the sensor at time $k$. The probability of detection at state $x \in S$ is denoted by $p_{\d,k}(x)$. The number of false alarms is Poisson distributed with parameter $\lambda_{\fa}$.

The PHD filter is parametrised according to the true model; however, one aspect of the possibilistic modelling is that it does not fully specify the dynamical behaviour of the targets or the errors in the observation model. Although this fact has limited consequences when using real data, it does affect performance assessment based on simulated data. In the Gaussian case, using the same expected value and covariance in the probabilistic and possibilistic models proved to be the most suitable. The equations of the possibilistic model considered for filtering are then $\uvx_k = G \uvx_{k-1} + \uvu_k$, with $\uvu_k$ described by $f_{\uvu}(u) = \oN(u; 0, Q)$ and we consider $\alpha_{\s} = 1$. The possibility function describing the initial velocity of targets is $\oN(0,\sigma_{\b}^2 I_2)$. We consider that $\uvy_k = H(\uvx_k - x_{s,k}) + \uvv_k$ with $\uvv_k$ described by $f_{\uvv}(v) = \oN(v; 0, \sigma'^2 I_2)$. The possibilities related to detection are assumed to be time-dependent by a straightforward generalisation of the model in Section~\ref{sec:observation} and we consider $\alpha_{\d}(x) = 1$ and $\alpha_{\df,k}(x) = 1 - p_{\d,k}(x)$. The parameters $\alpha_{\b,k}$ is assumed to be constant for $k > 1$ and is deduced from $\lambda_{\b}$ by multiplying the latter by $c = 2 \pi \sigma'^2 / V$, with $2\pi\sigma'^2$ the volume of the observation uncertainty, i.e.\ the integral of $f_{\uvv}$, and $V$ the volume of the observed area.

Pruning is used for both methods, with a threshold of $10^{-2}$ for the proposed method and of $5 \times 10^{-3}$ for the PHD filter, the difference stemming from the fact that possibilities tend to be larger than probabilities. Merging is also applied based on the Hellinger distance with a threshold of $0.1$ for the proposed method and with different merging criteria for the PHD filter: either with the Mahalanobis distance with a threshold of $4$ or with the Hellinger distance with a threshold of $0.1$. An analogue to the standard Hellinger distance which satisfies the same requirements is proposed in Appendix~\ref{sec:Hellinger}. As opposed to the Mahalanobis distance, the Hellinger distance is sufficiently conservative to be applied to the proposed approach. Track extraction is performed according to the method of Section~\ref{sec:track_extraction_1} with $\tau = 0.75$. The performance is assessed via the OSPA distance \cite{Schuhmacher2008}, which does not depend on the utilised representation of uncertainty and can therefore be used here without modifications. 

\subsection{Standard scenario}
\label{sec:standardScenario}

This first scenario is of duration $T = 25$ and has parameters $\sigma = 0.5$, $p_{\s} = 0.995$, $\sigma_{\b} = 5$, $\sigma' = 5$ and $\lambda_{\fa} = 10$.  We assume that the state of the sensor and the probability of detection are constant, i.e., for any $k \in \{1,\dots,T\}$, it holds that $x_{s,k} = \begin{bmatrix}0 & 0 & 0 & 0\end{bmatrix}^{\tr}$ and that $p_{\d,k}(x) = 0.9$ for any $x \in S_{\mathrm{obs}}$; it follows that $V$ is the volume of $S'$. False alarms are uniformly distributed on $S'$ and we consider $\alpha_{\fa} = c \lambda_{\fa}$. We assume that $\alpha_{\b,1} = \alpha_{\b,k}$ for any $k > 1$, i.e.\ the information provided to the proposed method at the first time step is equivalent to the one used in the PHD filter. For the PHD filter, each Gaussian component with a weight greater than $\tau_{\c} \in \{0.5, 0.75\}$ is considered as a track.

\begin{figure}
\centering
\includegraphics[width=.9\columnwidth, trim=127pt 320pt 140pt 330pt, clip]{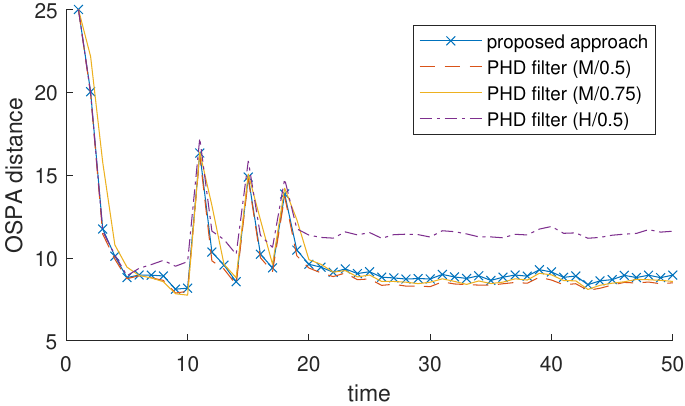}
\caption{OSPA distance with parameters $c=25$ and $p=2$ averaged over 1000 repeats of the observation process, where the GM-PHD filter uses two different merging criteria (H: Hellinger, M: Mahalanobis) and where $\tau_{\c} \in \{0.5, 0.75\}$.}
\label{fig:ospa}
\end{figure}

As seen in Figure~\ref{fig:ospa}, the performance of the PHD filter depends strongly on the choice of merging criteria. The proposed approach provides a compromise between the two by using the more conservative Hellinger-based merging while approaching the performance of the Mahalanobis-based PHD filter. The average computational time for a single time step in the PHD filter is $11.2\mathrm{ms}$ with Mahalanobis-based merging and $33.3\mathrm{ms}$ with Hellinger-based merging, and $31.4\mathrm{ms}$ for the proposed approach (with a Hellinger-based merging). Therefore, it appears that the computational overhead of the proposed approach stems mostly from the choice of merging criteria. In this scenario, the PHD filter has a better performance with $\tau_{\c} = 0.5$ than with $\tau_{\c} = 0.75$, the latter value making it less reactive to target birth. 

\subsection{Moving sensor}
\label{sec:movingSensor}

We consider a more challenging scenario where a moving sensor with limited field of view (FoV) monitors the space $S$ while being constrained in position to $S_{\s} = [100, 900] \times [100, 900] \subset S'$. The sensor's velocity is of a constant magnitude equal to $50$ and rotates by an angle of $\pm\pi/2$ when meeting the boundaries of $S_{\s}$. At each time step, the sensor's velocity vector is subject to a rotation by a normally-distributed random angle with mean $0$ and variance $0.01$.

This second scenario is of duration $T = 150$ and has parameters $\sigma = 0.01$, $p_{\s} = 0.999$, $\sigma_{\b} = 1$, $\sigma' = 5$ and $\lambda_{\fa} = 1$. The probability of detection is time-varying and modelled as $p_{\d,k}(x) = \oN(x; x_{\s,k}, \sigma_{\s}^2 I_2)$, with $\sigma_{\s} = 150$ modelling the extent of the sensor's FoV. False alarms are sampled from $\N(Hx_{\s,k}, \sigma_{\s}^2 I_2)$, so that $V = 2\pi\sigma_{\s}^2$ and we consider $F_{\fa}(y) = c\lambda_{\fa}\oN(y; H x_{\s,k}, \sigma_{\s}^2 I_2)$. When updating a term with expected value $\mu$, the probability of detection is assumed to be constant and equal to $p_{\d,k}(\mu)$, the possibility of detection is equal to $1$ everywhere and does not require approximation. The parameter $\alpha_{\b,1}$ is set to $1$, which means that the proposed approach has no information about the number of targets at the moment when the sensor is turned on; although this is important for multi-sensor applications, there is no analogue of this model in the probabilistic context.

In order to model the unseen targets, a 2-dimensional grid is defined in $S_{\mathrm{obs}}$ and the motion of these unseen targets is taken into account by convolving this grid with a Gaussian blur with variance $\sigma_{\b}^2I_2$, which corresponds to a random walk. Birth, survival and detection failures are then taken into account by point-wise operations following the equations of each approach. The main difference is that the birth intensity is initialised to $0$ in the PHD filter whereas $F_{\b}(x) = 1$ for any $x \in S$ is considered in the proposed approach; this means that the latter assumes no knowledge about the number of targets already present in the scene at the beginning of the scenario whereas the PHD filter relies on the fact that no targets are present at first.

\begin{figure}
\centering
\includegraphics[width=.9\columnwidth, trim=127pt 320pt 138pt 328pt, clip]{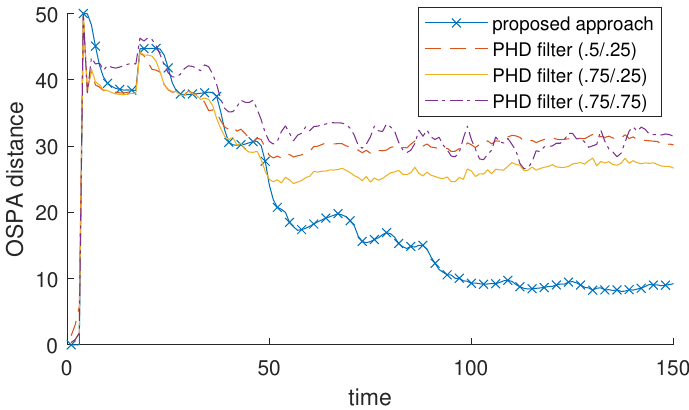}
\caption{OSPA distance with parameters $c=50$ and $p=2$ averaged over 1000 repeats of the observation process, where the GM-PHD filter is parametrised by $\tau_{\c} \in \{0.5, 0.75\}$ and $\tau_{\dc} \in \{0.25, 0.75\}$, indicated in the form $\tau_{\c} / \tau_{\dc}$.}
\label{fig:ospa_moving}
\end{figure}

Both approaches preserve the confirmed status of a track if detection failure has a larger posterior weight than all other data associations stemming from the same predicted track; this allows to keep tracks confirmed even when they are outside of the FoV of the sensor. The proposed approach performs two additional checks before confirming a track as follows:
\begin{enumerate*}
\item the effect of the initial uncertainty is tested with a threshold $\tau_{\b} = 0.1$; once a Gaussian term passes this test it remains \emph{pre-confirmed} and the test is no longer carried out, and 
\item the fitness of the predicted possibility function is assessed with a threshold $\tau_{\p} = \tau_{\b}$; this guarantees that associations with Gaussian terms that have not been detected for many time steps do not yield false tracks.
\end{enumerate*}
These tests are not required in the PHD filter due to the difference in behaviour between possibility functions and probability distributions; yet, they are not computationally burdensome and allow the proposed approach to operate with little prior knowledge on the number of unseen targets. In order to help the PHD filter maintain tracks, a de-confirmation threshold $\tau_{\dc}$ is implemented in such a way that a previously-confirmed track remains confirmed as long as the weight of the corresponding Gaussian component remains greater than $\tau_{\dc}$; this is shown to improve performance in Figure~\ref{fig:ospa_moving}.

When comparing the performance of the proposed approach with different parametrisations of the PHD filter in Figure~\ref{fig:ospa_moving}, it appears that the proposed method can better deal with the uncertainty in the number of unseen target in spite of the fact that, as opposed to the PHD filter, it does not assume any knowledge on target numbers at the initial time step. This added generality is crucial in applications where potentially many targets might already be present in the scene when the sensor is turned on. As opposed to the previous scenario, the PHD filter has better performance when $\tau_{\c} = 0.75$ when compared to $\tau_{\c} = 0.5$; the confirmation threshold $\tau$ for the proposed method has not been changed.

Other existing algorithms of higher computational complexity, such as the ones based on labelling strategies \cite{Vo2013, Reuter2014} or on mixed Poisson-Bernoulli representations \cite{Williams2015}, could largely outperform both the PHD filter and the proposed approach; yet, the PHD filter remains of importance because of its conceptual simplicity and its computational efficiency which transfer to its possibilistic analogue.

\section{Conclusion}
\label{sec:conclusion}

A variant of the notion of point process adapted to possibility theory was introduced and studied. This concept, referred to as uncertain counting measure, provides significant modelling versatility for multi-target systems, enabling for instance the representation of the absence of information about the number of targets and/or about their respective state. The notion of uncertain counting measure was then shown to lead to a recursion that is strikingly similar to the PHD filter, with sums and integrals replaced by maximums and supremums and with intensity functions replaced by presence functions. The properties and implementation of this recursion were discussed, followed by an assessment of its performance on simulated data.

Future work will aim to derive efficient algorithms based on the introduced model. Such algorithms could follow from introducing a labelling strategy \cite{Vo2013, Reuter2014}, from considering the associated smoothing problem \cite{Jiang2015} or from using a different representation of multi-target systems \cite{Houssineau2020} in order to introduce a track-based linear-complexity algorithm \cite{Houssineau2018_HISP}. By propagating more information, these algorithms could allow for tracking to be performed in the absence of information about the birth process at all time steps.

\bibliographystyle{abbrv}
\bibliography{Uncertainty}

\begin{thebibliography}{10}

\bibitem{BarShalom1990}
Y.~Bar-Shalom, T.~E. Fortmann, and P.~G. Cable.
\newblock Tracking and data association, 1990.

\bibitem{Baudrit2008}
C.~Baudrit, D.~Dubois, and N.~Perrot.
\newblock Representing parametric probabilistic models tainted with
  imprecision.
\newblock {\em Fuzzy sets and systems}, 159(15):1913--1928, 2008.

\bibitem{Berger1992}
J.~O. Berger and J.~M. Bernardo.
\newblock On the development of the reference prior method.
\newblock {\em Bayesian statistics}, 4(4):35--60, 1992.

\bibitem{Beynon2000}
M.~Beynon, B.~Curry, and P.~Morgan.
\newblock The {D}empster--{S}hafer theory of evidence: an alternative approach
  to multicriteria decision modelling.
\newblock {\em Omega}, 28(1):37--50, 2000.

\bibitem{Buede1997}
D.~M. Buede and P.~Girardi.
\newblock A target identification comparison of {B}ayesian and
  {D}empster-{S}hafer multisensor fusion.
\newblock {\em IEEE Transactions on Systems, Man, and Cybernetics-Part A:
  Systems and Humans}, 27(5):569--577, 1997.

\bibitem{Caron2011}
F.~Caron, P.~Del~Moral, A.~Doucet, and M.~Pace.
\newblock On the conditional distributions of spatial point processes.
\newblock {\em Advances in Applied Probability}, 43(2):301--307, 2011.

\bibitem{Chiu2013}
S.~N. Chiu, D.~Stoyan, W.~S. Kendall, and J.~Mecke.
\newblock {\em Stochastic geometry and its applications}.
\newblock John Wiley \& Sons, 2013.

\bibitem{Daley2003}
D.~J. Daley and D.~Vere-Jones.
\newblock {\em An introduction to the theory of point processes: volume I}.
\newblock Springer Science \& Business Media, 2003.

\bibitem{deBaets1999}
B.~De~Baets, E.~Tsiporkova, and R.~Mesiar.
\newblock Conditioning in possibility theory with strict order norms.
\newblock {\em Fuzzy Sets and Systems}, 106(2):221--229, 1999.

\bibitem{deMelo2018}
F.~E. De~Melo and S.~Maskell.
\newblock A {CPHD} approximation based on a discrete-{G}amma cardinality model.
\newblock {\em IEEE Transactions on Signal Processing}, 67(2):336--350, 2018.

\bibitem{DelMoral2015}
P.~Del~Moral and J.~Houssineau.
\newblock Particle association measures and multiple target tracking.
\newblock In {\em Theoretical Aspects of Spatial-Temporal Modeling}, pages
  1--30. Springer, 2015.

\bibitem{Delande2014}
E.~Delande, M.~{\"U}ney, J.~Houssineau, and D.~E. Clark.
\newblock Regional variance for multi-object filtering.
\newblock {\em IEEE Transactions on Signal Processing}, 62(13):3415--3428,
  2014.

\bibitem{Dempster1968}
A.~P. Dempster.
\newblock A generalization of {B}ayesian inference.
\newblock {\em Journal of the Royal Statistical Society: Series B},
  30(2):205--232, 1968.

\bibitem{Denoeux2000}
T.~Denoeux.
\newblock A neural network classifier based on dempster-shafer theory.
\newblock {\em IEEE Transactions on Systems, Man, and Cybernetics-Part A:
  Systems and Humans}, 30(2):131--150, 2000.

\bibitem{Denoeux2014}
T.~Denoeux, N.~El~Zoghby, V.~Cherfaoui, and A.~Jouglet.
\newblock Optimal object association in the {D}empster--{S}hafer framework.
\newblock {\em IEEE transactions on cybernetics}, 44(12):2521--2531, 2014.

\bibitem{Douc2009}
R.~Douc, E.~Moulines, Y.~Ritov, et~al.
\newblock Forgetting of the initial condition for the filter in general
  state-space hidden markov chain: a coupling approach.
\newblock {\em Electronic Journal of Probability}, 14:27--49, 2009.

\bibitem{Dubois2015}
D.~Dubois and H.~Prade.
\newblock Possibility theory and its applications: Where do we stand?
\newblock In {\em Springer Handbook of Computational Intelligence}, pages
  31--60. Springer, 2015.

\bibitem{Fortmann1980}
T.~E. Fortmann, Y.~Bar-Shalom, and M.~Scheffe.
\newblock Multi-target tracking using joint probabilistic data association.
\newblock In {\em 19th IEEE Conference on Decision and Control including the
  Symposium on Adaptive Processes}, pages 807--812, 1980.

\bibitem{Houssineau2018_parameter}
J.~Houssineau.
\newblock Parameter estimation with a class of outer probability measures.
\newblock {\em arXiv preprint arXiv:1801.00569}, 2018.

\bibitem{Houssineau2018}
J.~Houssineau and A.~Bishop.
\newblock Smoothing and filtering with a class of outer measures.
\newblock {\em SIAM/ASA Journal on Uncertainty Quantification}, 6(2):845--866,
  2018.

\bibitem{Houssineau2019}
J.~Houssineau, N.~Chada, and E.~Delande.
\newblock Elements of asymptotic theory with outer probability measures.
\newblock {\em arXiv preprint arXiv:1908.04331}, 2019.

\bibitem{Houssineau2018_HISP}
J.~Houssineau and D.~E. Clark.
\newblock Multitarget filtering with linearized complexity.
\newblock {\em IEEE Transactions on Signal Processing}, 66(18):4957--4970,
  2018.

\bibitem{Houssineau2020}
J.~Houssineau and D.~E. Clark.
\newblock On a representation of partially-distinguishable populations.
\newblock {\em Statistics}, 54(1):23--45, 2020.

\bibitem{Houssineau2010}
J.~Houssineau and D.~Laneuville.
\newblock {PHD} filter with diffuse spatial prior on the birth process with
  applications to {GM}-{PHD} filter.
\newblock In {\em 13th Conference on Information Fusion}, 2010.

\bibitem{Houssineau2017}
J.~Houssineau and B.~Ristic.
\newblock Sequential {M}onte {C}arlo algorithms for a class of outer measures.
\newblock {\em arXiv preprint arXiv:1708.06489}, 2017.

\bibitem{Jiang2015}
L.~Jiang, S.~S. Singh, and S.~Y{\i}ld{\i}r{\i}m.
\newblock Bayesian tracking and parameter learning for non-linear multiple
  target tracking models.
\newblock {\em IEEE Transactions on Signal Processing}, 63(21):5733--5745,
  2015.

\bibitem{Li2018_joint}
T.~Li, H.~Chen, S.~Sun, and J.~M. Corchado.
\newblock Joint smoothing and tracking based on continuous-time target
  trajectory function fitting.
\newblock {\em IEEE Transactions on Automation Science and Engineering},
  16(3):1476--1483, 2018.

\bibitem{Mahler2003}
R.~P.~S. Mahler.
\newblock Multitarget {B}ayes filtering via first-order multitarget moments.
\newblock {\em IEEE Transactions on Aerospace and Electronic systems},
  39(4):1152--1178, 2003.

\bibitem{Mahler2007}
R.~P.~S. Mahler.
\newblock {\em Statistical Multisource-Multitarget Information Fusion}.
\newblock Artech House, 2007.

\bibitem{Mori1986}
S.~Mori, C.-Y. Chong, E.~Tse, and R.~Wishner.
\newblock Tracking and classifying multiple targets without a priori
  identification.
\newblock {\em IEEE Transactions on Automatic Control}, 31(5):401--409, 1986.

\bibitem{Nagappa2011}
S.~Nagappa and D.~E. Clark.
\newblock On the ordering of the sensors in the iterated-corrector probability
  hypothesis density ({PHD}) filter.
\newblock In {\em Signal Processing, Sensor Fusion, and Target Recognition XX},
  volume 8050, page 80500M, 2011.

\bibitem{Nam2016}
H.~Nam and B.~Han.
\newblock Learning multi-domain convolutional neural networks for visual
  tracking.
\newblock In {\em Proceedings of the IEEE conference on computer vision and
  pattern recognition}, pages 4293--4302, 2016.

\bibitem{Oussalah2000}
M.~Oussalah and J.~De~Schutter.
\newblock Possibilistic {K}alman filtering for radar {2D} tracking.
\newblock {\em Information Sciences}, 130(1-4):85--107, 2000.

\bibitem{Powell2006}
G.~Powell, D.~Marshall, P.~Smets, B.~Ristic, and S.~Maskell.
\newblock Joint tracking and classification of airbourne objects using particle
  filters and the continuous transferable belief model.
\newblock In {\em 9th IEEE Conference on Information Fusion}, 2006.

\bibitem{Reid1979}
D.~Reid.
\newblock An algorithm for tracking multiple targets.
\newblock {\em IEEE transactions on Automatic Control}, 24(6):843--854, 1979.

\bibitem{Reuter2014}
S.~Reuter, B.-T. Vo, B.-N. Vo, and K.~Dietmayer.
\newblock The labeled multi-{B}ernoulli filter.
\newblock {\em IEEE Transactions on Signal Processing}, 62(12):3246--3260,
  2014.

\bibitem{Ristic2010}
B.~Risti\'c, D.~Clark, and B.-N. Vo.
\newblock Improved {SMC} implementation of the {PHD} filter.
\newblock In {\em 13th Conference on Information Fusion}, 2010.

\bibitem{Ristic2012}
B.~Ristic, D.~Clark, B.-N. Vo, and B.-T. Vo.
\newblock Adaptive target birth intensity for {PHD} and {CPHD} filters.
\newblock {\em IEEE Transactions on Aerospace and Electronic Systems},
  48(2):1656--1668, 2012.

\bibitem{Ristic2018}
B.~Ristic, J.~Houssineau, and S.~Arulampalam.
\newblock Robust target motion analysis using the possibility particle filter.
\newblock {\em IET Radar, Sonar \& Navigation}, 13(1):18--22, 2018.

\bibitem{Ristic2019}
B.~Ristic, J.~Houssineau, and S.~Arulampalam.
\newblock Target tracking in the framework of possibility theory: The
  possibilistic {B}ernoulli filter.
\newblock {\em Information Fusion}, 62:81--88, 2020.

\bibitem{Salmond1990}
D.~J. Salmond.
\newblock Mixture reduction algorithms for target tracking in clutter.
\newblock In {\em SPIE signal and data processing of small targets}, volume
  1305, pages 434--445, 1990.

\bibitem{Schlangen2017}
I.~Schlangen, E.~D. Delande, J.~Houssineau, and D.~E. Clark.
\newblock A second-order {PHD} filter with mean and variance in target number.
\newblock {\em IEEE Transactions on Signal Processing}, 66(1):48--63, 2017.

\bibitem{Schuhmacher2008}
D.~Schuhmacher, B.-T. Vo, and B.-N. Vo.
\newblock A consistent metric for performance evaluation of multi-object
  filters.
\newblock {\em IEEE transactions on signal processing}, 56(8):3447--3457, 2008.

\bibitem{Shafer1976}
G.~Shafer.
\newblock {\em A mathematical theory of evidence}, volume~42.
\newblock Princeton university press, 1976.

\bibitem{Singh2009}
S.~S. Singh, B.-N. Vo, A.~Baddeley, and S.~Zuyev.
\newblock Filters for spatial point processes.
\newblock {\em SIAM Journal on Control and Optimization}, 48(4):2275--2295,
  2009.

\bibitem{Stone2013}
L.~D. Stone, R.~L. Streit, T.~L. Corwin, and K.~L. Bell.
\newblock {\em Bayesian multiple target tracking}.
\newblock Artech House, 2013.

\bibitem{Streit2010}
R.~L. Streit.
\newblock {\em Poisson point processes: imaging, tracking, and sensing}.
\newblock Springer Science \& Business Media, 2010.

\bibitem{Vo2006}
B.-N. Vo and W.-K. Ma.
\newblock The {G}aussian mixture probability hypothesis density filter.
\newblock {\em IEEE Transactions on Signal Processing}, 54(11):4091--4104,
  2006.

\bibitem{Vo2005}
B.-N. Vo, S.~Singh, and A.~Doucet.
\newblock Sequential {M}onte {C}arlo methods for multitarget filtering with
  random finite sets.
\newblock {\em IEEE Transactions on Aerospace and electronic systems},
  41(4):1224--1245, 2005.

\bibitem{Vo2013}
B.-T. Vo and B.-N. Vo.
\newblock Labeled random finite sets and multi-object conjugate priors.
\newblock {\em IEEE Transactions on Signal Processing}, 61(13):3460--3475,
  2013.

\bibitem{Washburn1987}
R.~B. Washburn.
\newblock A random point process approach to multiobject tracking.
\newblock In {\em IEEE American Control Conference}, pages 1846--1852, 1987.

\bibitem{Williams2015}
J.~L. Williams.
\newblock Marginal multi-bernoulli filters: Rfs derivation of mht, jipda, and
  association-based member.
\newblock {\em IEEE Transactions on Aerospace and Electronic Systems},
  51(3):1664--1687, 2015.

\bibitem{Zadeh1978}
L.~A. Zadeh.
\newblock Fuzzy sets as a basis for a theory of possibility.
\newblock {\em Fuzzy sets and systems}, 1(1):3--28, 1978.

\end{thebibliography}

\appendices

\section{Hellinger distance for possibility functions}
\label{sec:Hellinger}

The Hellinger distance $\mathrm{H}(p, q)$ between two probability density functions $p$ and $q$ defined on the same set $S \subseteq \bbR^d$ is characterised by
\begin{equation}
\label{eq:probaHellinger}
\mathrm{H}^2(p, q) = \dfrac{1}{2} \int \big( \sqrt{p(x)} - \sqrt{q(x)} \big)^2 \d x,
\end{equation}
and verifies $\mathrm{H}(p, q) \in [0,1]$. The Hellinger distance takes a value of $1$ when $p$ and $q$ have disjoint supports, in which case the integral in \eqref{eq:probaHellinger} simplifies to $\int p(x) \d x + \int q(x) \d x = 2$. The coefficient $1/2$ can then be seen as a normalising constant.

Now considering two possibility functions $f$ and $g$ defined on $S$, a natural analogue of \eqref{eq:probaHellinger} can be introduced as
$$
\overline{\mathrm{H}}^2(f, g) 
= \dfrac{\int \big( \sqrt{f(x)} - \sqrt{g(x)} \big)^2 \d x}{\int f(x) \d x + \int g(x) \d x} 
$$
The function $\overline{\mathrm{H}}(\cdot, \cdot)$ is a distance and verifies $\overline{\mathrm{H}}^2(f, g) \in [0,1]$. In the special case where $f(x) = \overline{\mathrm{N}}(x; \mu_1, P_1)$ and $g(x) = \overline{\mathrm{N}}(x; \mu_2, P_2)$, it holds that
\begin{multline*}
\overline{\mathrm{H}}^2(f, g) = 1 - \dfrac{2\sqrt{|P_1||P_2|}}{\sqrt{|P|} \big(\sqrt{|P_1|} + \sqrt{|P_2|}\big)} \\
\times \exp\Big( -\dfrac{1}{8} (\mu_1 - \mu_2)^{\tr} P^{-1} (\mu_1 - \mu_2) \Big),
\end{multline*}
with $P = (P_1 + P_2)/2$ and with $|\cdot|$ denoting the determinant.

\section{Pseudo-code}
\label{sec:pseudoCode}

The pseudo-code for the implementation of the proposed approach based on a Gaussian max-mixture is given in Algorithm~\ref{alg:GMPHD}. In order to accommodate for the uncertainty in location at birth, the information filter is used in the update instead of the usual Kalman filter recursion. This allows for infinite variance to be taken into account formally. We denote by $\mu_{\b}$ and $I_{\b}$ the expected value and precision at birth and assume that the first observation is informative enough to make the precision after the first update positive-definite. The standard Kalman filter could be used with a spatially informative birth, in which case the variance $P_{\b}$ would be related to the precision $I_{\b}$ via $P_{\b} = I_{\b}^{-1}$.

\begin{algorithm} 
\caption{Gaussian max-mixture implementation}
\label{alg:GMPHD}
\begin{flushleft}
\textbf{Input:} Indexed set $\{(\hat{w}_{k-1,i}, \hat{\mu}_{k-1,i}, \hat{P}_{k-1,i})\}_{i = 1}^{n_{k-1}}$ and observation set $Y_k = \{y_{k,1}, \dots, y_{k,m_k}\}$ \\
\end{flushleft}
\begin{algorithmic}
\ForAll{$i \in \{1, \dots, n_{k-1}\}$} \Comment{Prediction}
\State $w_{k,i} \leftarrow \alpha_{\s} \hat{w}_{k-1,i}$
\State $\mu_{k,i} \leftarrow G \hat{\mu}_{k-1,i}$
\State $P_{k,i} \leftarrow G \hat{P}_{k-1,i} G^{\tr} + Q$
\State $I_{k,i} \leftarrow P_{k,i}^{-1}$ \Comment{Convert to precision}
\EndFor
\State $i \leftarrow  n_{k-1}+1$ \Comment{Birth}
\State $w_{k,i} \leftarrow \alpha_{\b,k}$
\State $\mu_{k,i} \leftarrow \mu_{\b}$
\State $I_{k,i} \leftarrow I_{\b}$
\State $r_j \leftarrow 0$ \Comment{Initialise denominator}
\ForAll{$i \in \{1,\dots, n_{k-1}+1\}$} \Comment{Update}
\State $\hat{y}_{k,i} \leftarrow H \mu_{k,i}$ \Comment{Predicted observation}
\State $S_{k,i} \leftarrow H P_{k,i} H^{\tr} + R$ \Comment{Covariance of innovation}
\ForAll{$j \in \{1,\dots,m_k\}$}
\State $l \leftarrow (i-1)m_k + j$
\State $\tilde{w}_{k,l} \leftarrow \alpha_{\d} w_{k,i} \overline{\mathrm{N}}(y_{k,j}; \hat{y}_{k,i}, S_{k,i})$
\State $\hat{P}_{k,l} \leftarrow (I_{k,i} + H^{\tr} R^{-1} H)^{-1}$ 
\State $\hat{\mu}_{k,l} \leftarrow \hat{P}_{k,l} ( I_{k,i}\mu_{k,i} + H^{\tr} R^{-1} y_{k,j} )$ 
\State $r_j \leftarrow \max\{r_j, \tilde{w}_{k,l} \}$
\EndFor
\EndFor
\ForAll{$i \in \{1,\dots, n_{k-1}+1\}$} \Comment{Weight normalisation}
\ForAll{$j \in \{1,\dots,m_k\}$}
\State $l \leftarrow (i-1)m_k + j$
\State $\hat{w}_{k,l} \leftarrow \tilde{w}_{k,l} / r_j$
\EndFor
\EndFor
\ForAll{$i \in \{1,\dots, n_{k-1}\}$} \Comment{Detection failure}
\State $l \leftarrow (n_{k-1}+1)m_k + i$
\State $\hat{w}_{k,l} \leftarrow \alpha_{\df} w_{k,i}$
\State $\hat{\mu}_{k,l} \leftarrow \mu_{k,i}$
\State $\hat{P}_{k,l} \leftarrow P_{k,i}$
\EndFor
\State $n_k \leftarrow (n_{k-1}+1)m_k + n_{k-1}$
\end{algorithmic}
\begin{flushleft}
\textbf{Output:} Indexed set $\{(\hat{w}_{k,i}, \hat{\mu}_{k,i}, \hat{P}_{k,i})\}_{i = 1}^{n_k}$
\end{flushleft}
\end{algorithm}

In practice, pruning and merging must be applied to the output $\{(\hat{w}_{k,i}, \hat{\mu}_{k,i}, \hat{P}_{k,i})\}_{i = 1}^{n_k}$ of Algorithm~\ref{alg:GMPHD}; the only difference is that the weight of a Gaussian term after merging is the maximum of the weights of all the different merged components. 

\section{Multivariate Gaussian possibility function}

Consider a vector $\mu \in \bbR^n$ and a $n \times n$ positive definite matrix $P$, then the multivariate Gaussian possibility function with expected value $\mu$ and covariance matrix $P$ is defined as
$$
\N(x; \mu, P) = \exp\Big( -\dfrac{1}{2}(x - \mu)^{\tr}P^{-1}(x - \mu) \Big), \qquad x \in \bbR^n.
$$
One can indeed check that is $\uvx$ is described by $\N(\mu, P)$ then $\bbE^*(\uvx) = \mu$ and $\bbV^*(\uvx) = P$. 

Now consider that $\uvx$ and $\mu$ are of the form $\uvx = [\uvx_1^{\tr},\uvx_2^{\tr}]^{\tr}$ and $\mu = [\mu_1^{\tr},\mu_2^{\tr}]^{\tr}$ and $P$ is of the form
$$
P =
\begin{bmatrix}
P_1 & P_{1,2} \\
P_{2,1} & P_2
\end{bmatrix},
$$
with $\uvx_1$ and $\mu_1$ of dimension $p < n$ and $P_1$ of dimension $p \times p$. Standard linear algebra results yield
\begin{multline*}
(x - \mu)^{\tr}P^{-1}(x - \mu) = (x_1 - \mu_1)^{\tr}P_1^{-1}(x_1 - \mu_1) \\
+ (x_2 - \mu_{2|1}(x_1))^{\tr}P_{2|1}^{-1}(x_2 - \mu_{2|1}(x_1))
\end{multline*}
with $\mu_{2|1}(x_1) = \mu_2 + P_{2,1} P_1^{-1}(x_1 - \mu_1)$ and $P_{2|1} = P_2 - P_{2,1} P_1^{-1}P_{1,2}$. The marginal possibility function describing $\uvx_1$ can then be deduced as
$$
f_{\uvx_1}(x_1) = \sup_{x_2 \in \bbR^q} \N\big([x_1^{\tr},x_2^{\tr}]^{\tr}; \mu, P\big) = \N(x_1; \mu_1, P_1)
$$
with $q = n - p$. Indeed, the supremum is reached at $x_2 = b(x_1)$ from which the result follows easily. The conditional possibility function describing $\uvx_2$ given $\uvx_1 = x_1$ can also be deduced as
$$
f_{\uvx_2|\uvx_1}(x_2 \given x_1) = \dfrac{\N\big([x_1^{\tr},x_2^{\tr}]^{\tr}; \mu, P\big)}{\N(x_1; \mu_1, P_1)} = \N(x_2; \mu_{2|1}(x_1), P_{2|1}).
$$

\section{Spatial necessity w.r.t.\ the initial uncertainty}
\label{sec:init_spatialnecessity}

\subsection{Definition}

When there is a significant initial uncertainty, we aim to assess the fitness of the prior possibility function with respect to an observation $y \in Y_k$ for a Gaussian term created at time $k_1$ associated with the observations $y_{k_1}, \dots, y_{k_m}$ at the respective times $k_1,\dots,k_m$ with $k_1 < \dots < k_m < k$. If we define $h_{k|k_1}(\cdot \given x_{k_1}, y_{k_{1:m}})$ as the possibility function describing the observation at time $k$ given the state $x_{k_1}$ at time $k_1$ and the observations $y_{k_1}, \dots, y_{k_m}$ then the fitness of the prior $f_{\uvx_{k_1}}$ is defined as
$$
\Lambda_{k|k_1}(y ; y_{k_{1:m}}) = 1 - \sup_{x \in \sfX} (1 - h_{k|k_1}(y \given x, y_{k_{1:m}})) f_{\uvx_{k_1}}(x \given y_{k_{1:m}}).
$$
One can check that the corresponding upper bound is indeed equal to the marginal likelihood $\Gamma_k(y; y_{k_{1:m}}) = \sup_{x \in X} h(y \given x) f_{\uvx_k}(x \given y_{k_{1:m}})$. Under mild conditions, the conditional possibility function $h_{k|k_1}(\cdot \given x_{k_1}, y_{k_{1:m}})$ will be less and less dependent on $x_{k_1}$ because of the forgetting properties of the filter (as in the probabilistic case \cite{Douc2009}) so that $\Lambda_{k|k_1}(y ; y_{k_{1:m}})$ will tend to $\Gamma_k(y; y_{k_{1:m}})$ as $k$ increases. The additional condition for track extraction can then be formulated as
$$
\dfrac{\Lambda_{k|k_1}(y ; y_{k_{1:m}})}{\Gamma_k(y; y_{k_{1:m}})} > \tau_{\b},
$$
for some fixed threshold $\tau_{\b} \in (0,1)$. This condition allows to distinguish between terms that have a large marginal likelihood because of the lack of information in the prior and those which accurately predict the next observation. There is no direct equivalent of this test in the probabilistic context for the same reasons as the ones behind the existence of Bayes factors: the probabilistic marginal likelihood is difficult to interpret on its on. The computation of $\Lambda_{k|k_1}(y ; y_{k_{1:m}})$ in the linear-Gaussian case is considered in the next section.

\subsection{Computation}

The objective in this section is to compute the spatial necessity of the observation $y \in Y_{k+1}$ given the previous observations $y_1, \dots, y_k$, at time steps $1$ to $k$, in the linear-Gaussian case. The spatial necessity of interest can be expressed as
\begin{equation}
\label{eq:necessity}
\Lambda_{k+1}(y ; y_{1:k}) = 1 - \sup_{x \in \sfX} (1 - h_{k+1|1}(y \given x, y_{1:k})) f_{\uvx_1}(x \given y_{k_{1:k}}),
\end{equation}
where $h_{k+1|1}(\cdot \given x, y_{1:k})$ is the possibility function describing the observation at time $k+1$ given the state at time $1$ and the observations $y_1,\dots,y_k$. Cases with missing observations can be treated similarly. To simplify the presentation, we assume that the prior possibility function $f_{\uvx_1}$ is of the form $\N(\mu_{\b}, P_{\b})$ with the covariance matrix $P_{\b}$ having finite elements. We first express the prior possibility function describing the joint state $\uvx_{1:k} = [\uvx_1^{\tr},\dots,\uvx_k^{\tr}]^{\tr}$ as $f_{\uvx_{1:k}}(x_{1:k}) = \N(x_{1:k}; \mu_{1:k}, P_{1:k})$ with
$$
\mu_{1:k} =
\begin{bmatrix}
\mu_{\b} \\ G \mu_b \\ \dots \\ G^{k-1} \mu_{\b}
\end{bmatrix}
$$
and
$$
P_{1:k} =
\begin{bmatrix}
P_1 & P_1 G^{\tr}  & \dots & P_1 (G^{\tr})^{k-1} \\
G P_1 & P_2 &  \dots & P_2 (G^{\tr})^{k-2} \\
\vdots   & \vdots    & \ddots         & \vdots \\
G^{k-1} P_1 & G^{k-2} P_2 & \dots & P_k
\end{bmatrix},
$$
where the covariance matrices $P_1, \dots, P_k$ are defined recursively as $P_{i+1} = GP_iG^{\tr} + Q$ for any $i \in \{1,\dots,k-1\}$ with $P_1 = P_{\b}$. Similarly, the extended observation vector is defined as $y_{1:k} = [y_1^{\tr},\dots,y_k^{\tr}]^{\tr}$ and the extended observation matrix $H_{1:k}$ is defined as the $d'k \times dk$ matrix of the form
$$
H_{1:k} =
\begin{bmatrix}
H & \bm{0}_{d',d} & \dots & \bm{0}_{d',d} \\
\bm{0}_{d',d} & H & \dots & \bm{0}_{d',d} \\
\vdots & & \ddots & \vdots \\
\bm{0}_{d',d} & \dots & & H
\end{bmatrix},
$$
where $\bm{0}_{d',d}$ is the null matrix of size $d' \times d$. In the remainder of this section, we will use a hat to indicate that quantities are conditioned on $y_{1:k}$. The Kalman filter can be used to compute the posterior possibility function $\hat{f}_{\uvx_{1:k}}(x_{1:k}) = \N(x_{1:k}; \hat{\mu}_{1:k}, \hat{P}_{1:k})$, from which the expected value $\hat{\mu}_1$ and variance $\hat{P}_1$ associated with the smoothing possibility function $\hat{f}_{\uvx_1}$ can be recovered as usual. The posterior variance $\hat{P}_{k|1}$ associated with $\hat{f}_{\uvx_k|\uvx_1}(\cdot \given x_1)$ does not depend on the initial state $x_1$ and can be expressed as $\hat{P}_{k|1} = \hat{P}_{k} - \hat{P}_{k,1} \hat{P}_1^{-1}\hat{P}_{1,k}$, where $\hat{P}_{i,j}$ denotes the $(i,j)$th block of $\hat{P}_{1:k}$ and $\hat{P}_i = \hat{P}_{i,i}$. The corresponding expected value $\hat{\mu}_{k|1}(x_1)$ does depend on $x_1$ and can be expressed in a linear form as $\hat{\mu}_{k|1}(x_1) = \hat{P}_{k,1} \hat{P}_1^{-1} x_1 + (\hat{\mu}_k - \hat{P}_{k,1} \hat{P}_1^{-1} \hat{\mu}_1)$. The conditional possibility function $h_{k+1|1}(y \given x, y_{1:k})$ can then be expressed as
\begin{multline*}
h_{k+1|1}(y \given x_1, y_{1:k}) = \\
\N\big( y; HG\hat{\mu}_{k|1}(x_1), H(G\hat{P}_{k|1}G^{\tr} + Q)H^{\tr} + R \big)
\end{multline*}

For long scenarios, computing the full smoothing possibility function $\hat{f}_{\uvx_{1:k}}$ might become impractical. However, this step becomes unnecessary as soon as the spatial necessity becomes close enough to the marginal likelihood $\Gamma_{k+1}(y; y_{1:k})$ so that these calculations only need to be performed over the initial period of existence of the corresponding Gaussian mixture term, which varies depending on the variance of the noise in the transition and in the observation process.

Unfortunately, the maximisation in \eqref{eq:necessity} does not seem to have an analytical solution and numerical methods must be used. One simple approximation can be obtained by sampling $N$ times from a normal p.d.f.\ with mean $\hat{\mu}_{1|k}$ and variance $\hat{P}_{1|k}$ and by using the obtained random grid to evaluate the supremum in \eqref{eq:necessity}. The simulations presented in the article are based on $N = 1000$ samples.

\section{Proofs}

\subsection{Proof of \Cref{res:outerIntensityMeasure}}

Let $\calX = \sum_{i=1}^{\uvn} \delta_{\uvx_i}$ be an uncertain counting measure. We want to prove that
$$
\bar{F}_{\calX}(B) = \bar\bbE\Big( \max_{i \in \{1,\dots, \uvn\}} \ind{B}(\uvx_i) \Big) = \bar{\bbP}(\calX(B) > 0),
$$
for any $B \subseteq \sfX$. We first express the argument of $\bar{\bbE}(\bscdot)$ as
$$
\max_{i \in \{1,\dots, \uvn\}} \ind{B}(\uvx_i) = \ind{B_{\uvn}}(\uvx_1,\dots,\uvx_n) = \bar{\bbI}_{(\uvx_1,\dots,\uvx_n) \in B_{\uvn}}
$$
where the subset $B_{\uvn}$ of $\sfX^{\uvn}$ is defined as
\eqns{
B_{\uvn} = \bigcup_{i=1}^{\uvn} \sfX \times \dots \times \sfX \times \underbrace{B}_{\mathclap{\text{$i$\textsuperscript{th} position}}} \times \sfX \times \dots \times \sfX,
}
and where $\bar{\bbI}_E$ is the indicator of the event $E$ in $\Omega$. By construction, it holds that $\bar{\bbE}(\bar{\bbI}_E) = \bar{\bbP}(E)$ for any event $E$, so that
$$
\bar{F}_{\calX}(B) = \bar{\bbP}( (\uvx_1,\dots,\uvx_n) \in B_{\uvn} ).
$$
We conclude the proof by identifying the event $(\uvx_1,\dots,\uvx_n) \in B_{\uvn}$ with the event $\sum_{i=1}^{\uvn} \ind{B}(\uvx_i) > 0$ and by noticing that $\sum_{i=1}^{\uvn} \ind{B}(\uvx_i) = \calX(B)$.

\subsection{Proof of \Cref{thm:sumIndUcm}}

The possibility function $f_{\calZ}$ describing the uncertain counting measure $\calZ = \calX+\calX'$ can be expressed as
\eqnsml{
f_{\calZ}(z_1,\dots,z_m) = \\
\max_{\substack{n,k : n+k=m  \\ \sigma \in \Sym(m)}} f_{\calX}\big(z_{\sigma(1)},\dots,z_{\sigma(n)}\big) f_{\calX'}\big(z_{\sigma(n+1)},\dots,z_{\sigma(n+k)} \big)
}
for any $(z_1,\dots,z_m) \in \sfX^m$ and any $m \in \bbN_0$, where $f_{\calX}$ and $f_{\calX'}$ are describing $\calX$ and $\calX'$ respectively and where $\Sym(m)$ is the set of permutations of $\{1,\dots,m\}$. It follows that
\eqnsa{
F_{\calZ}(z) & = \max_{m:m>0}\bigg(\sup_{z_{2:m} \in \sfX^m} f_{\calZ}(z, z_2, \dots, z_m) \bigg)\\
& = \max_{n:n>0} \bigg(\sup_{x_{2:n} \in \sfX^n} f_{\calX}(z, x_2, \dots, x_n) \bigg)\\
& \qquad\qquad \lor \max_{k : k>0} \bigg(\sup_{x_{2:k} \in \sfX^k} f_{\calX'}(z, x_2, \dots, x_k)\bigg) \\
& = \max \{ F_{\calX}(z), F_{\calX'}(z) \}.
}

\subsection{Proof of \Cref{thm:predictionFirstMoment}}

The possibility function describing the uncertain counting measure $\calZ$ given $\calX = \sum_{i=1}^{\uvn} \delta_{\uvx_i}$ takes the form
\eqns{
f_{\calZ}(z_1,\dots,z_m \given \calX) = \ind{\uvn}(m) \max_{\sigma \in \Sym(m)} \prod_{i=1}^m g(z_i \given \uvx_i).
}
Noticing that the number of points is not affected by the prediction through $g(\bscdot \given x)$, we conclude that
\eqnsa{
F_{\calZ}(z_1) & = \sup_{(z_2,\dots,z_m) \in \bar{\bbZ}} f_{\calZ}(z_{1:m}) \\
& = \max_{m > 0} \Bigg(\sup_{\substack{z_{2:m} \in \sfZ^{m-1} \\ x_{1:m} \in \sfX^m}} f_{\calX}(x_{1:m} \given m) \prod_{i=1}^m g(z_i \given x_i) \Bigg) \\
& = \sup_{x_1 \in \sfX} \bigg( g(z_1 \given x_1) \sup_{x_{2:n} \in \bar{\bbX}} f_{\calX}(x_{1:n}) \bigg) \\
& = \sup_{x_1 \in \sfX} g(z_1 \given x_1) F_{\calX}(x_1).
}

\subsection{Proof of \Cref{thm:updatedIntensity}}
\label{proof:updatedIntensity}

We consider the extended state space $\bar{\sfX}$ introduced in \Cref{rem:extension} and the corresponding presence function and likelihood function. We also introduce $\calX_{\fa,k}$ as the uncertain counting measure $\uvn_{\fa,k} \delta_{\psi_{\fa}}$ with $\uvn_{\fa,k}$ the unknown number of false-alarm generators at time $k$. This type of modelling cannot be used with the corresponding sets since sets cannot represent multiplicity, e.g.\ $\{a, a\} = \{a\}$ for any element $a$. We then denote $\bar\calX_k$ the uncertain counting measure resulting from the superposition of $\calX_k$ and $\calX_{\fa,k}$, i.e.\ $\bar\calX_k = \calX_k + \calX_{\fa}$. Since there is no information about $\uvn_{\fa,k}$, it indeed holds that $F_k(\psi_{\fa} \given Y_{1:k-1}) = 1$. In the remainder of the proof, we condition implicitly on $Y_{1:k-1}$ and write, e.g., $F_k$ instead of $F_k(\bscdot \given Y_{1:k-1})$ or $F_k(\bscdot \given Y_k)$ instead of $F_k(\bscdot \given Y_{1:k})$. The points in $\bar\calX_k$ are i.i.d.\ by the possibility function $f_k = F_k$ since the presence function $\bar{F}_k$ is a possibility function and therefore $\ninf{F_k} = 1$. When conditioning on a possibly-empty observation $y \in \sfY$, we obtain the posterior possibility function
\begin{multline*}
f_k(x_k \given y) = \ind{\phi}(y) h(\phi \given x_k)f_k(x_k) \\
\lor \ind{S'}(y) \dfrac{h(y \given x_k)f_k(x_k)}{ \sup_{x \in \bar{\sfX}} h(y \given x)f_k(x)},
\end{multline*}
for any $x \in \bar{\sfX}$. Defining $\uvm_{\phi}$ as the uncertain number of detection failures at time $k$, we introduce the superposition
$$
\bar{\calY}_k = \sum_{i=1}^{\bar{\uvm}} \delta_{\bar{\uvy}_i} = \calY_k + \uvm_{\phi} \delta_{\phi}
$$
of $\calY_k$ with the uncertain counting measure $\uvm_{\phi} \delta_{\phi}$ representing detection failures. We omit the time subscripts in the points of $\calY_k$ as well, i.e.\ $\calY_k = \sum_{i=1}^{\uvm} \delta_{\uvy_i}$. The possibility function describing $\bar{\calX}_k$ given $\bar{\calY}_k$ is
$$
f_{\bar{\calX}_k}(x_1,\dots,x_n \given \bar\calY_k) = \ind{\bar{\uvm}}(n) \max_{\sigma \in \Sym(n)} \prod_{i=1}^n f_k(x_{\sigma(i)} \given \bar{\uvy}_i),
$$
for any $x_1,\dots,x_n \in \bar{\sfX}$ and any $n \geq 0$. Since there is no available information about $\uvm_{\phi}$, it holds that
\begin{align*}
& f_{\bar{\calY}_k}(\bar{y}_1,\dots,\bar{y}_m \given \calY_k) = \ind{[\uvm,\infty)}(n) \\
& \times \max_{\sigma \in \Sym(m)} \ind{\uvy_{1:\uvm}}(\bar{y}_{\sigma(1)},\dots,\bar{y}_{\sigma(\uvm)}) \prod_{i = \uvm+1}^m \ind{\phi}(\bar{y}_{\sigma(i)}),
\end{align*}
that is, points in $\bar{\calY}_k$ must match with points in $\calY_k$, except for those at $\phi$. An expression for the possibility function describing $\bar{\calX}_k$ given $\calY_k$ can then be obtained as
\begin{multline*}
f_{\bar{\calX}_k}(x_{1:n} \given \calY_k) = \ind{[\uvm,\infty)}(n) \\
\times \max_{\sigma \in \Sym(n)} \prod_{i=1}^{\uvm} f_k(x_{\sigma(i)} \given \uvy_i) \prod_{i=\uvm+1}^n f_k(x_{\sigma(i)} \given \phi).
\end{multline*}
The posterior possibility function describing $\calX_k$ can be deduced from the one describing $\bar{\calX}_k$ as
$$
f_{\calX_k}(x_{1:n} \given \calY_k) = \max_{n_{\fa} \geq 0} f_{\bar{\calX}_k}(x_1, \dots, x_n, \psi_{\fa}, \dots, \psi_{\fa} \given \calY_k),
$$
for any $x_1,\dots,x_n \in \sfX$ and any $n \geq 0$, since, intuitively, the uncertain counting measure $\calX_k$ does not specify the number of points at $\psi_{\fa}$ that $\bar{\calX}_k$ contains. The proof is concluded by computing the posterior presence function given $\calY_k$ as
$$
F_k(x_k \given \calY_k) = f_{\phi}(1) f_k(x_k \given \phi)\lor f_{\phi}(0) \max_{1 \leq i \leq \uvm} f_k(x_k \given \uvy_i),
$$
for any $x_k \in \sfX$, which follows from the fact that $f_{\bar{\calX}_k}(\bscdot \given \calY_k)$ is maximised when the number $m$ of detection failures is minimised, i.e.\ $m = 0$ in the case of detection or $m = 1$ in the case of detection failure. The posterior presence function can then be expressed more explicitly as
\begin{align*}
F_k(x_k \given \calY_k) & = h(\phi \given x_k) F_k(x_k) \\
& \quad \lor \max_{1 \leq i \leq \uvm} \dfrac{h(\uvy_i \given x_k)F_k(x_k)}{F_{\fa}(\uvy_i) \lor \sup_{x \in \sfX} h(\uvy_i \given x)F_k(x)},
\end{align*}
as desired.

\end{document}